\documentclass[twocolumn,aps,prb,amsmath,amssymb,longbibliography,superscriptaddress]{revtex4}
\usepackage[english]{babel}
\usepackage{graphicx,bm,amssymb,amsmath}
\usepackage{color,wrapfig,braket}
\usepackage{hyperref,placeins,ulem}
\usepackage[caption=false]{subfig}

\usepackage{subfig}

\usepackage{epsfig}
\usepackage{verbatim}
\usepackage{yhmath}
\def\S{\textrm{S}}

\makeatletter
\newcommand\xleftrightarrow[2][]{%
  \ext@arrow 9999{\longleftrightarrowfill@}{#1}{#2}}
\newcommand\longleftrightarrowfill@{%
  \arrowfill@\leftarrow\relbar\rightarrow}
\makeatother

\begin{document}

\title{The Existence of Featureless Paramagnets on the Square and the Honeycomb Lattices in 2+1D}
\author{Chao-Ming Jian}
\affiliation{Department of Physics, Stanford University, Stanford, California 94305, USA}
\affiliation{Kavli Institute for Theoretical Physics, University of California, Santa Barbara, CA 93106, USA}
\author{Michael Zaletel}
\affiliation{Station Q, Microsoft Research, Santa Barbara, CA 93106, USA}
\date{\today}

\begin{abstract}
%The central idea of the Landau paradigm is to characterize continuous phase transitions between different phases by the spontaneous breaking of the their underlying symmetries. An hidden assumption of this paradigm is the existence of a fully symmetric or featureless phase. In fact, this assumption is sometimes false at zero temperature. Certain featureless states that are gapped and are not classically or topologically ordered have already been proven to be impossible by the Lieb-Schultz-Mattis theorem and its generalizations. In general, it is important to give a definite answer to the question whether or not a featureless phase exists under given conditions either by a proof of its non-existence or by an example. In this paper, we focus on the spin-1 square lattice and the spin-1 and spin-1/2 honeycomb lattice with spin rotation, translation and other point group symmetries in 2+1D. Although featureless paramagnet phases in these cases are not ruled out by any theorem, field theoretic arguments strongly disfavor their existence. Nevertheless, as a proof of principle, we provide examples of them using a slave-spin PEPS construction. The featureless-ness of the spin-1 slave-spin PEPS wave functions on the square lattice and the honeycomb lattice are verified both analytically and numerically. A proposal for the spin-1/2 featureless paramagnet on the honeycomb lattice is provided.

The peculiar features of quantum magnetism sometimes forbid the existence of gapped `featureless' paramagnets which are fully symmetric and unfractionalized.
The Lieb-Schultz-Mattis theorem is an example of such a constraint, but it is not known what the most general restriction might be.
We focus on the existence of featureless paramagnets on the spin-1 square lattice and the spin-1 and spin-1/2 honeycomb lattice with spin rotation and space group symmetries in 2+1D. Although featureless paramagnet phases are not ruled out by any existing theorem, field theoretic arguments disfavor their existence. Nevertheless, by generalizing the construction of  Affleck, Kennedy, Lieb and Tasaki to a class we call `slave-spin' states, we propose featureless wave functions for these models. The featureless-ness of the spin-1 slave-spin states on the square and honeycomb lattice are verified both analytically and numerically, but the status of the spin-1/2 honeycomb state remains unclear.

\end{abstract}

\maketitle
\section{Introduction}
In the Landau paradigm   phases of matter are classified by their symmetries, and fully symmetric phases are ubiquitous. Only after spontaneous breaking of these symmetries do classical orders emerge. One  reason  the Landau paradigm is incomplete is that a gapped and fully symmetric phase is sometime excluded in a zero temperature quantum phase diagram.
%An implicit assumption of the Landau paradigm of phase transitions is that a fully symmetric phase does exist. At finite temperature this assumption is justified, since the symmetry of the system is always restored at high temperature. However, for quantum phases at zero temperature, the existence of a gapped and fully symmetric phase is not  guaranteed. 
For example, the Lieb-Schultz-Mattis (LSM) theorem \cite{Lieb1961} asserts that an S=1/2 spin chain in 1+1D does not admit a gapped paramagnetic phase that respects all the symmetries,  specifically translation and $SO(3)$ spin rotations; the system is either gapless or  breaks a symmetry.
A generalized version of LSM theorem in higher dimensions  finds that a magnet with half-integer spin per unit cell cannot be a short-range entangled paramagnet; the system is either gapless, breaks a symmetry, or has fractionalized excitations (topological order). \cite{Hastings2004, Oshikawa}
%Translated from the spin language to particle language, the LSM theorem is equivalent to the statement that a short-range entangled fully symmetric Mott insulator does not exist for boson/fermion systems at half filling. 
We refer to a gapped, symmetric and short-range entangled paramagnet as a `featureless' paramagnet. The LSM theorem only requires translation and spin rotational symmetries, though versions incorporating additional space-group symmetries are also possible. \cite{Parameswaran2013, Watanabe2015}

%So far, the LSM theorem only concerns about translation symmetry and the global spin rotation symmetry. By extending the overall symmetry group, one might expect further constraints on the existence of a featureless insulator (or paramagnet) even with integer of particle  (or certain integral spin) per unit cell\cite{Parameswaran2013, Watanabe2015}. For example, Ref. \onlinecite{Parameswaran2013} shows that, on 3D lattices with non-symmorphic lattice symmetries, a charge conserved featureless insulator or a at-least $U(1)$ spin rotation invariant featureless paramagnet does not exist unless the total charge or spin per unit cell is a multiple of the non-symmorphic rank $s$ (157 of 230 space groups in 3D has $s>1$). In general, other than these cases in which theorems are already proved, it is still a very interesting question that whether or not a lattice with a fixed total spin/charge per unit cell and certain global symmetries can admit a featureless paramagnet/insulator phase.

To determine whether a featureless phase can exist in a  model (as defined by its Hilbert space and symmetries - not a Hamiltonian), one must either prove an extension of the LSM theorem forbidding its existence, or produce an explicit example.
Here we  focus on the spin-1 square lattice and the spin-1 and spin-1/2 honeycomb lattices in 2+1D.
Although not forbidden by any existing theorems, it is unknown whether featureless phases exist for these models.
In fact, field theoretic arguments based on the non-linear sigma model find non-trivial spatially dependent  Berry phase patterns \citep{Haldane1988, ReadSachdev1990, Senthil2004} which strongly disfavor the existence of such featureless phases,  at least  proximate to Neel or valence-bond orders where the field theory is justified.

Finding featureless paramagnetic wave functions is non-trivial as they require a finite amount of quantum entanglement.
For example, Ref. \onlinecite{Kimchi2013}  first explored this issue for the spin-1/2 honeycomb model at zero magnetization, requiring only $\textrm{U}(1)$ symmetry, and found that only strongly interacting states can provide an example of a featureless paramagnet.\cite{BraydenBauer} 
Requiring full $\textrm{SO}(3)$ symmetry further complicates the search.
In this work, we propose and analyze a class of featureless $\textrm{SO}(3)$-symmetric wave functions for the square and honeycomb magnets, and find conclusive evidence for the existence of featureless spin-1 paramagnets on the square and honeycomb lattices. Our numerics are unable to establish whether the proposed spin-1/2 honeycomb state has exponentially decaying correlations.

%In general, even the construction of a featureless insulator with $U(1)$ symmetry might encounter some complication, the construction of a featureless paramagnet with $SO(3)$ spin rotation symmetry (plus other lattice symmetry) can be even more non-trivial. 

%In many cases with global $U(1)$ and other lattice symmetries, non-interacting band insulators already provide good examples of featureless insulating states with integer particle number per unit cell. However, this is not always the case. 

The wave functions we propose are written in terms of projected entangled pair states (PEPS) \cite{Verstraete2004}. In general, global symmetries impose strong constraints on the PEPS\cite{García2010,Weichselbaum2012,Singh2012,Jiang2015, Singh2011,Bauer2011,Singh2013,Williamson2014}. To serve our purpose, we choose the `entangled pairs' to be multiplets of SO(3) symmetric spins and, thus, will refer to them as the `slave-spin' PEPS. By construction these states are fully symmetric, though numerical evidence is required to prove that they are gapped states  without spontaneously broken symmetry (``cat states") or topological order. These slave-spin PEPS suggest a generalized version of the deconfined quantum criticality when brought to a second order phase transition to a classically-ordered phase.

The rest of the paper is organized as follows. In  Sec. \ref{Spin_Chain} we discuss the 1D spin-1 chain as a warm up, in order to demonstrate in a simple setting how field theoretic arguments based on non-trivial Berry phases (the topological $\Theta$-term) can be evaded through the slave-spin PEPS construction. 
In Sec. \ref{Square} we study the spin-1 square lattice. After reviewing the field theoretic obstacle \cite{Haldane1988,ReadSachdev1990} to a featureless state, 
 we construct a featureless wave function with spin rotation, translation,  $C_4$ lattice rotations, and time reversal symmetry.
A numerical analysis verifies its featureless-ness.
In Sec. \ref{Honeycomb}, we study the spin-1 and spin-1/2 honeycomb lattices. 
We derive a similar field theoretic obstacle for these models, and construct a wave function with $SO(3)$ spin rotation, translation,  $C_3$ lattice rotations, mirror, and  time reversal symmetry.
For the spin-1 magnet, our numerics unambiguously verify its featureless-ness, while for spin-1/2 it is unclear whether the proposed state is gapless, or simply has a large correlation length we cannot yet detect.
In Sec. \ref{Conclusion} we conclude with a discussion of open questions and future directions.

\section{1+1D Featureless Paramagnets on the Spin-1 Chain}
\label{Spin_Chain}

We warm up with a 1+1D spin chain with spin rotation and translation symmetry, both to review the field theoretic Berry-phase analysis and introduce the slave-spin formalism in a simple setting.
%In this section, we will focus on the featureless paramagnetic phases of the 1+1D spin-1 chain as a warm up. A featureless paramagnetic phase in 1+1D is defined to be a short-range-entangled gapped phase with the lattice translation and the $SO(3)$ spin symmetry. First, we will start from the field theory analysis and explain a seemingly universal expectation on a featureless paramagnetic phase based on the contribution of a non-trivial Berry phase term (the $\Theta$-term) in the Lagrangian. 
%
%Secondly, we will review the AKLT state \cite{AKLT} as a featureless paramagnet that agrees with the field theoretic expectation, whose wave function is directly obtained from a slave-spin PEPS construction. Afterwards, we will present a generalized slave-spin PEPS which is also a featureless paramagnet and, more interestingly, lies beyond the field theory expectation.
%\subsection{Field Theoretic Analysis for a Featureless Paramagnet}
%In the field theory, we often describe a paramagnet phase as the disorder phase of a classical order parameter (the Neel order parameter in this case).
The field theory proceeds by deriving an effective action for the Neel order parameter  $\mathbf{n}\in S^2$.\cite{Haldane1983}
For a spin chain with spin $S$ per site, in addition to the kinetic terms there is a non-trivial topological Berry phase  (the $\Theta$-term)\cite{Altland_Simons2010}:
\begin{align}
S_{\text{top}}[\mathbf{n}]=\frac{\Theta}{4\pi} \int dx d\tau \mathbf{n} \cdot(\partial_x \mathbf{n} \times \partial_\tau \mathbf{n}),
\label{ThetaTerm}
\end{align}
where $\Theta=2\pi \S$. For integer spin  there is a gap in the bulk, leading to a featureless gapped paramagnet, the Haldane phase.\cite{Haldane1983}
For $S=1$ the term $S_{\text{top}}[\mathbf{n}]$ nevertheless leads to interesting degenerate edge states for an open chain. At $\S=1$,  $S_{\text{top}}$ is a total derivative in the bulk, but induces a $O(3)$ Wess-Zumino-Witten (WZW) term at level 1 on the boundary, which implies a spin-$1/2$ degree of freedom is localized on the boundary of the spin chain\cite{Ng1994}.
%This paramagnetic phase is exactly the Haldane phase\cite{Haldane1983} in 1+1D with its boundary spin-1/2 degree of freedom protected by the global $SO(3)$ spin rotation symmetry\cite{Chen2011}.

%\subsection{AKLT State in the Slave-Spin PEPS Form: An Ideal Wave Function in the Haldane Phase}

\begin{figure}[tb]
\centerline{
\includegraphics[width=3
in]{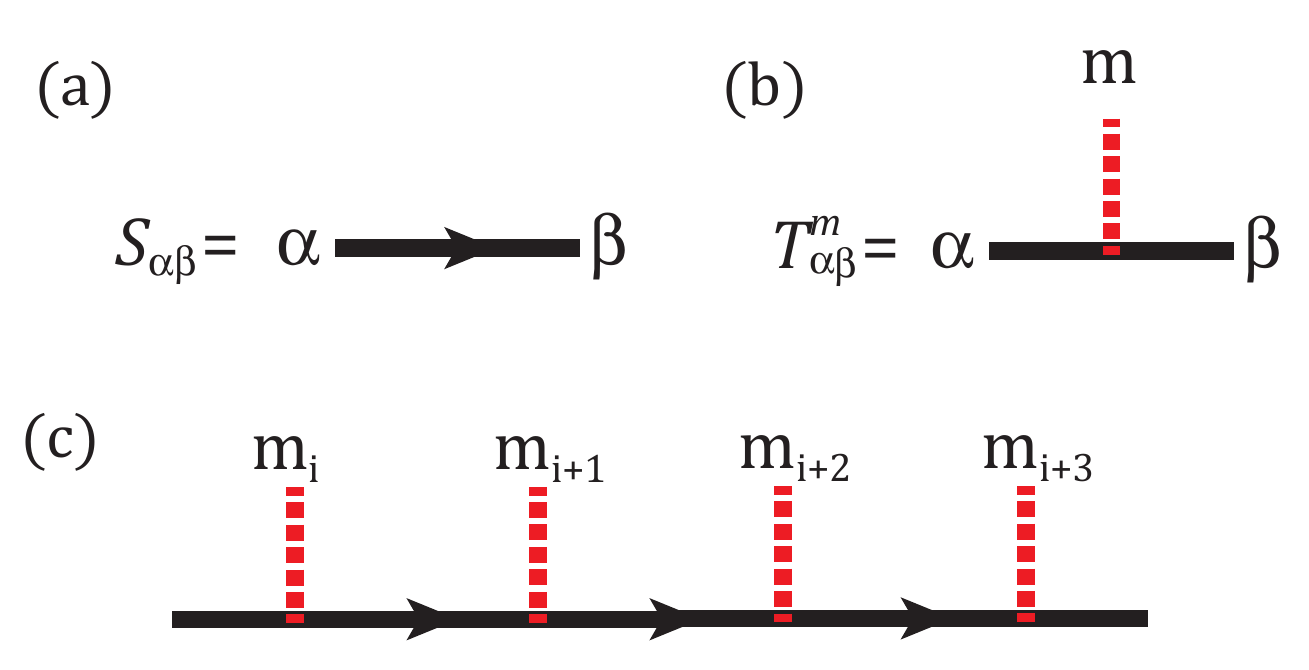}
}
\caption{\label{1D_SpinH_Tensor} (a) $S_{\alpha\beta}$ represents the projector of two spin-1/2 degrees of freedom onto the spin singlet channel. The arrow indicates the anti-symmetric property of $S_{\alpha\beta}$. (b) $T^m_{\alpha\beta}$ with symmetric subscripts represents the projector of two spin-1/2 degrees of freedom onto the spin triplet channel. (c) The graph represents the tensor network the AKLT wave function. The red lines represent physical spin-1 degrees of freedom, whereas the joined black lines represent the auxiliary spin-1/2 degrees of freedom that are contracted.}
\end{figure}

An ideal wave function for the spin-1 Haldane phase and its edge states is given by the Affleck-Kennedy-Lieb-Tasaki (AKLT) state, \cite{AKLT} the simplest example of a `slave-spin' state.
% In the following, we review its construction in the form of the matrix product state (MPS) using the slave-spin PEPS language. The AKLT state in the standard MPS form can be written as
%\begin{align}
%|\Psi_\text{AKLT}\rangle =\sum_{\{m\}} \left(\prod_i A^{m_i}\right) |...m_i m_{i+1}...\rangle
%\end{align}
%where $m_i=1,2,3$ represent the three spin-1 states with $\hat{S}_z=1,0,-1$ on the $i$th site ($\hat{S}_z$ is the $z$ component of the spin operator) and $A^m$s are $2\times2$ matrices given by
%\begin{align}
%A^1=  \sqrt{\frac{2}{3}}&\left( \begin{array}{cc}
%0 & 1 \\ 0 & 0
%\end{array}\right),~
%A^2=\frac{1}{\sqrt{3}}\left(\begin{array}{cc}
%1 & 0 \\ 0 & -1
%\end{array}\right), \nonumber \\  
%& A^3=\sqrt{\frac{2}{3}}\left(\begin{array}{cc}
%0 & 0 \\ -1 & 0
%\end{array}\right).
%\label{AKLTTensor1}
%\end{align}
%Now, we review the derivation of this wave function using slave-spin PEPS language, which is easily generalizable to other 1D chains and to higher dimensional spin models.
We start by associating two auxiliary spin-1/2  degrees of freedom labeled by $|\alpha_i \beta_i\rangle$ on each site $i$,  with $\alpha_i, \beta_i = \uparrow / \downarrow$.
A featureless state $|\Psi_0\rangle$ for the auxiliary spins is obtained by projecting the $|\beta_i\rangle$ on the $i$th site and $|\alpha_{i+1}\rangle$ on the $(i+1)$th site onto the spin-0 channel:
\begin{align}
|\Psi_0\rangle=\sum_{\{\alpha,\beta\}} \left(\prod_i S_{\beta_{i},\alpha_{i+1}}\right) |...\alpha_{i}\beta_{i}\alpha_{i+1}\beta_{i+1}...\rangle, 
\label{Singlet_Matrix}
\end{align}
with $S$ a $2\times 2$ matrix:
\begin{align}
S=\left(
\begin{array}{cc}
0 & 1 \\ -1 & 0
\end{array}
\right).
\label{CG_HHto0}
\end{align}
 $S$ is obtained from the Clebsh-Gordon (CG) coefficient of ``fusing" two spin-1/2s into a spin singlet, namely $S_{\alpha\beta}\equiv\langle \S=0| \S=1/2,\alpha;\S=1/2,\beta\rangle$.
 To obtain a wave function for the physical spin-1 degrees of freedom, we then project the two auxiliary spin-1/2s onto a spin-1 channel of each site:
\begin{align}
& \Psi_\text{AKLT}(\{m_i\})  =\langle...m_im_{i+1}...|\Psi\rangle_0 \nonumber \\
& = \sum_{\{\alpha,\beta\}} \left(\prod_i T^{m_i}_{\alpha_{i},\beta_{i}}S_{\beta_{i},\alpha_{i+1}}\right),
\label{AKLTwavefunction}
\end{align}
where $m_i=1,2,3$ represent the three spin-1 states on the $i$th site. The tensor $T^{m}_{\alpha,\beta} \equiv \langle \S=1, m| \S=1/2,\alpha;\S=1/2,\beta\rangle$ is the CG coefficient for fusing two spin-1/2s into a spin triplet.
% whose is explicitly form is given by
%\begin{align}
%T^1=  \left( \begin{array}{cc}
%1 & 0 \\ 0 & 0
%\end{array}\right),~
%T^2=\frac{1}{\sqrt{2}}\left(\begin{array}{cc}
%0 & 1 \\ 1 & 0
%\end{array}\right),   
% T^3=\left(\begin{array}{cc}
%0 & 0 \\ 0 & 1
%\end{array}\right).
%\label{CG_HHto1}
%\end{align}
Graphical representations of the tensors $S_{\alpha\beta}$ and $T^m_{\alpha\beta}$  and their contraction are shown in Fig. \ref{1D_SpinH_Tensor}. 
The wave function of  Eq. \ref{AKLTwavefunction} is actually in the form of a matrix product state (MPS), as it takes the form of a matrix-trace in the auxiliary $\alpha / \beta$ space.
The `bond dimension' $\chi$ of the MPS is  $\chi=2$, corresponding to two the auxiliary spin states. 
For a finite open spin chain each boundary has one un-summed matrix index that corresponds to  the  spin-1/2 edge state of the Haldane phase, consistent with the field theory prediction.

%\subsection{Exotic 1D Featureless Paramagnet from the Slave-Spin PEPS Construction}

\begin{figure}[tb]
\centerline{
\includegraphics[width=3
in]{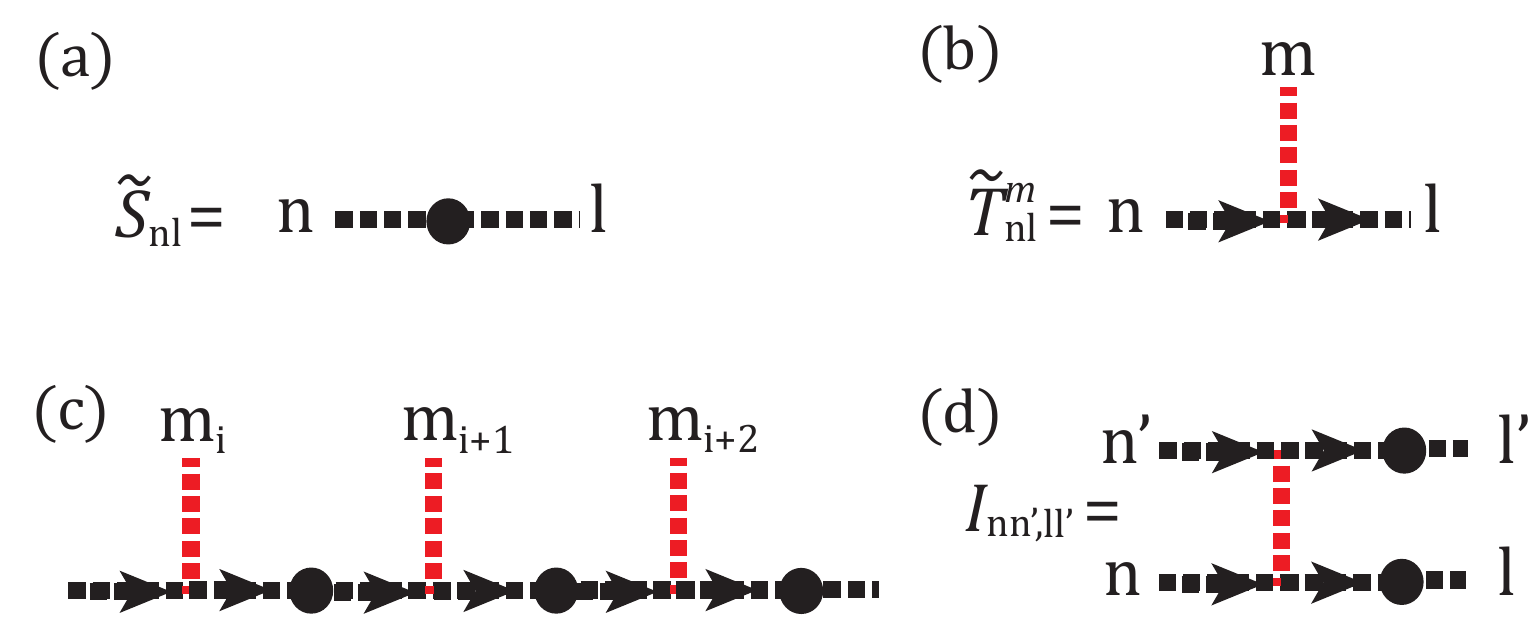}
}
\caption{\label{1D_Spin1_Tensor} 
(a) $\tilde{S}_{\alpha\beta}$ is a symmetric matrix that represents the projector of two spin-1 degrees of freedom onto the total-spin-singlet channel. (b) $\tilde{T}^m_{\alpha\beta}$ with anti-symmetric subscripts represents the projector of two spin-1 degrees onto the spin-1 channel. (c) The graph represents tensor network for the wave function $|\Psi_\text{para}\rangle$. The red lines represent physical spin-1 degrees of freedom, whereas the joined black lines represent the auxiliary spin-1 degrees of freedom that are contracted.
}
\end{figure}

Is the boundary spin-1/2 inevitable in any paramagnetic phase of the spin-1 anti-ferromagnet?
%Now, we are interested in the following question: Does the field theoretic prediction on the boundary degrees of freedom based on the topological Berry phase term $S_{\text{top}}$ dictate all possible featureless paramagnet in a spin-1 chain? 
No: it is simple to construct a slave-spin counterexample to  this field theoretic expectation.
Instead of the auxiliary spin-1/2s of the AKLT case, we start with two auxiliary spin-1s on each site $i$, with the states $|l_i n_i\rangle,  l_i, n_i \in 1, 2, 3$ labeling the $\hat{S}_z$ basis. We again project pairs $|n_i\rangle$ and $|l_{i+1}\rangle$ (for all $i$) into the spin-0 channel by using the CG coefficient $\tilde{S}_{n_i l_{i+1}}\equiv\langle \S=0| \S=1,n_i; \S=1,l_{i+1}\rangle$.
%\begin{align}
%\tilde{S}=\frac{1}{\sqrt{3}}\left(
%\begin{array}{ccc}
%0 & 0 & 1 \\ 0 &-1 & 0 \\ 1 & 0 & 0
%\end{array}
%\right).
%\label{CG_11to0}
%\end{align}
We then project the auxilliary spins of each site $|l_i n_i\rangle$ onto the total spin-1 channel in order to obtain a spin-1 chain wave function. This projection is accomplished by the CG coefficients $\tilde{T}^{m}_{l_i, n_i} \equiv \langle \S=1, m| \S=1, l_i ;\S=1, n_i\rangle$.
%:
%\begin{align}
%\tilde{T}^1=  \frac{1}{\sqrt{2}} &\left( \begin{array}{ccc}
%0 & 1 & 0 \\ -1 & 0 & 0 \\ 0 & 0 & 0
%\end{array}\right),~
%\tilde{T}^2=\frac{1}{\sqrt{2}}\left(\begin{array}{ccc}
%0 & 0 & 1 \\ 0 & 0 & 0 \\ -1 & 0 & 0
%\end{array}\right),  \nonumber \\ 
%& ~~~~~\tilde{T}^3=\frac{1}{\sqrt{2}}\left(\begin{array}{ccc}
%0 & 0 & 0 \\ 0 & 0 & 1 \\ 0 & -1 & 0
%\end{array}\right).
%\label{CG_11to1}
%\end{align}
The new spin-1 slave-spin  wave function is
\begin{align}
|\Psi_\text{para}\rangle =\sum_{\{m\}} \left(\prod_i \tilde{T}^{m_i} \tilde{S}\right) |...m_i m_{i+1}...\rangle.
\end{align}
The graphical representation of the tensors $\tilde{S}$ and $\tilde{T}$ is given in Fig. \ref{1D_Spin1_Tensor}.
% (a) and (b) with the arrow in Fig. \ref{1D_Spin1_Tensor} (b) representing the anti-symmetry of the two subscripts of $\tilde{T}^m_{nl}$. The construction of the wave function $|\Psi_\text{para}\rangle$ is given by the tensor network shown in Fig. \ref{1D_Spin1_Tensor} (c) (red lines for physical degrees of freedom and black lines for auxiliary degrees of freedom). 
Since the state is composed by successive \textrm{SO}(3) invariant projections,  by construction it is invariant under global spin rotations and is translation invariant. 
However, we must verify it is not a `cat-state' of symmetry breaking valence-bond patterns.
The standard MPS test for such states is to analyze the `transfer matrix' (Fig. \ref{1D_Spin1_Tensor} (d)) defined  by 
\begin{align}
I_{nn',ll'}=\sum_m \left(\tilde{T}^{m} \tilde{S} \right)^*_{n'l'} \left(\tilde{T}^{m} \tilde{S} \right)_{nl}.
\end{align}
If the dominant eigenvector of $I_{nn',ll'}$ is unique, which we find it is,  all equal-time correlation decay exponentially, so $|\Psi_\text{para}\rangle$ is a featureless paramagnet.

Since we used spin-1 auxiliary degrees of freedom, on an open chain the edge state of $|\Psi_\text{para}\rangle$ are effectively spin-1, counter to the field theoretic expectation.
%Therefore, if the field theoretic description is given by $S_{\text{top}}[\mathbf{n}]$ in Eq. \ref{ThetaTerm} with effective $\Theta_\text{AKLT}=2\pi$ for the fractionalized spin-1/2 singlet pairs, the effective $\Theta$ that describes the fractionalized spin-1 singlet pairs for $|\Psi_\text{para}\rangle$ should be $\Theta_\text{para}=2\Theta_\text{AKLT}=4\pi$, which was unexpected in the seemingly universal field theoretic analysis. 
 In fact, $|\Psi_\text{para}\rangle$ represents the ``trivial phase" in the $\mathbf{Z}_2$ classification of $SO(3)$ spin symmetry protected topological states, while the AKLT  state represents the ``non-trivial phase"\cite{Pollmann, Fidkowski, Chen2011}.
 
In conclusion, while the seemingly universal field theoretic analysis is highly suggestive (the ground state of the nearest-neighbor Heisenberg chain is indeed in the Haldane phase) it does not `mandate' any properties of the energy spectrum in the same sense that the Lieb-Schultz-Mattis theorem does.
%The two are separated by a phase transition if the $SO(3)$ spin symmetry is always respected.

%An intuitive argument for the difference between the two states $|\Psi_\text{AKLT}\rangle$ and $|\Psi_\text{para}\rangle$ is the following. The slave-spin PEPS construction is essentially a method to construct a projective wave function using fractionalized spin degrees of freedom. The on-site spin-1 degrees of freedom are fractionalized into two spin 1/2s in $|\Psi_\text{AKLT}\rangle$ and into two spin 1s $|\Psi_\text{para}\rangle$. In both cases, the fractionalized degrees of freedom form singlet pairs between every pair of neighboring sites. Therefore, if the field theoretic description is given by $S_{\text{top}}[\mathbf{n}]$ in Eq. \ref{ThetaTerm} with effective $\Theta_\text{AKLT}=2\pi$ for the fractionalized spin-1/2 singlet pairs, the effective $\Theta$ that describes the fractionalized spin-1 singlet pairs for $|\Psi_\text{para}\rangle$ should be $\Theta_\text{para}=2\Theta_\text{AKLT}=4\pi$, which was unexpected in the seemingly universal field theoretic analysis. 
%

\section{2+1D Spin-1 featureless  paramagnet on the square lattice}
\label{Square}
In this section, we study the featureless paramagnetic phase on the square lattice with spin 1 per site in 2+1D under the following symmetries: the global $SO(3)$ spin rotation, the translation, the lattice $C_4$ rotation ($R_{\frac{2\pi}{4}}$) the time reversal ($\mathcal{T}$) and the mirror symmetries ($\sigma_\text{v}$ with the vertical mirror plane and $\sigma_{45^\circ }$ with the $45^\circ$ mirror plane). We will first review the field theoretic obstacle to such a spin-1 featureless paramagnetic phase. Then, we prove its existence by providing an example wave function using the slave-spin PEPS construction. A numerical analysis verifies its featureless-ness. In the end, we will comment on the connection to previous studies Ref. \onlinecite{Jiang2009,Wang2015} on the spin-1 square lattice.

\subsection{Field Theoretic Obstacle to Featureless Paramagnetic Phases on the Square Lattice}
\label{Square_Field_Theory}
\begin{figure}[tb]
\centerline{
\includegraphics[width=3.5
in]{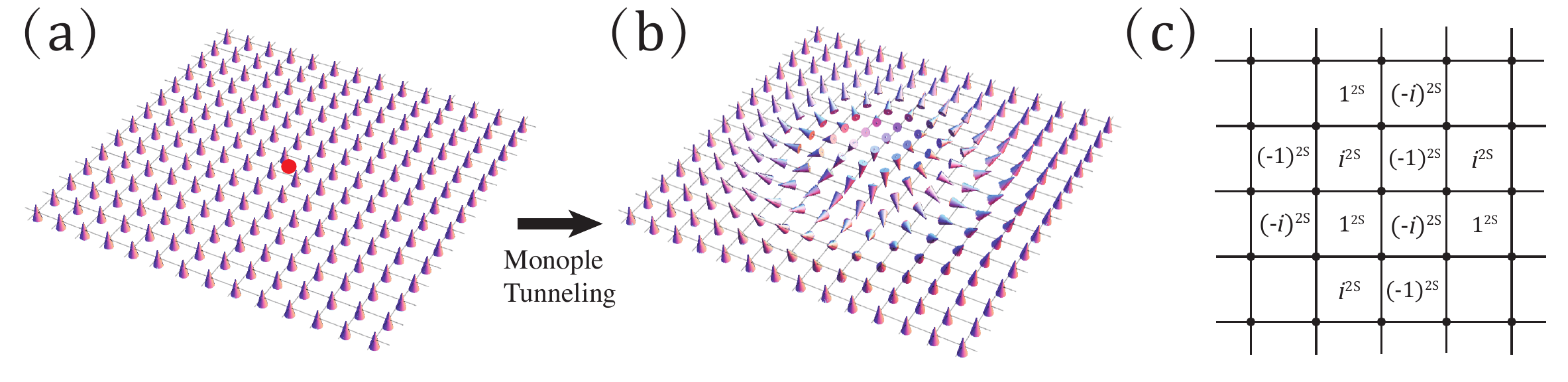}
}
\caption{\label{Square_Phase_Pattern} A spatially uniform configuration of the Neel order parameter with a trivial winding number (a) tunnels into a configuration in a different topological sector with winding number 1 after a monopole tunneling event occurred at the plaquette center. (c) shows the spatial dependence of the non-trivial Berry phase associated to the monopole tunneling event in a spin-$\S$ Neel background. }
\end{figure}

From a field theory point of view, to obtain a paramagnetic phase, one can start from an classically ordered background and disorder it through the proliferation of topological defects/excitations or instantons. 
In the Neel background in 2+1D, the topological instanton to be considered is the monopole tunneling event\cite{Haldane1988} which happens at plaquette centers and changes the winding number (or the homotopy class) of the spatial configuration of the Neel order parameter at a given time (see Fig. \ref{Square_Phase_Pattern} (a) and (b)). There are non-trivial location-dependent Berry phase associated to the monopole tunneling events in the spin-coherent-state path integral \cite{Haldane1988, ReadSachdev1990}. With spin $\S$ per site, this Berry phase is $(+1)^{2\S}$, $i^{2\S}$, $(-1)^{2\S}$ or $(-i)^{2\S}$ depending on whether the dual lattice coordinate $(x,y)$ of plaquette center where the monopole tunneling event happens is (even,even), (even, odd), (odd, odd) or (odd, even) (see Fig. \ref{Square_Phase_Pattern} (c)).

For $\S=1/2$, this non-trivial Berry phase leads to the deconfined quantum criticality\cite{Senthil2004}. When the monopole tunneling event is proliferated in the spacetime and the Neel background is disordered, the non-trivial location dependence of this Berry phase in Fig. \ref{Square_Phase_Pattern} (c) leaves an imprint, either  breaking a space group symmetry or  resulting in topological order. Therefore, a featureless paramagnetic phase does not exist for $\S=1/2$ on the square lattice, which agrees with the generalized LSM theorem\cite{Hastings2004}. Similar results applies for $\S=3/2$. For $\S=1$, even though the generalized LSM theorem does not apply, the Berry phase still has a non-trivial spatial pattern which results in either classical orders, a nematic phase\cite{ReadSachdev1990} for instance, or topological states\cite{Freedman2004} when monopole tunneling events are proliferated. The obstacle to a featureless paramagnetic phase remains for $\S=1$. Furthermore, the Berry phase pattern is trivial when $\S=2$, in which case an example of featureless paramagnets is given by the 2D AKLT states on the square lattice.

\subsection{Spin-1 Featureless Paramagnets on the Square Lattice}
\label{Square_FP}
\begin{figure}[tb]
\centerline{
\includegraphics[width=1.6
in]{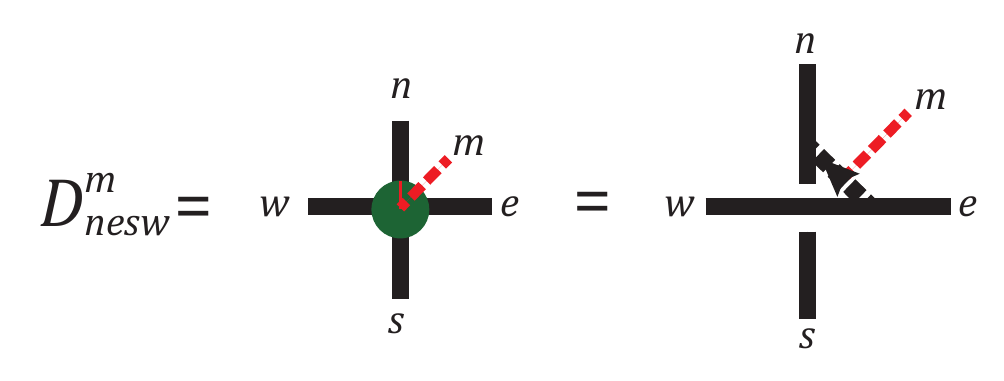}
}
\caption{\label{Square_SiteTensor1} The graphical representation of the tensor $D^m_{nesw}$ that implements the projection of 4 spin-1/2 degrees of freedom to a total-spin-1 channel compatible with $R_\frac{2\pi}{4}$ and $\mathcal{T}$.}
\end{figure}

\begin{figure}[tb]
\centerline{
\includegraphics[width=3
in]{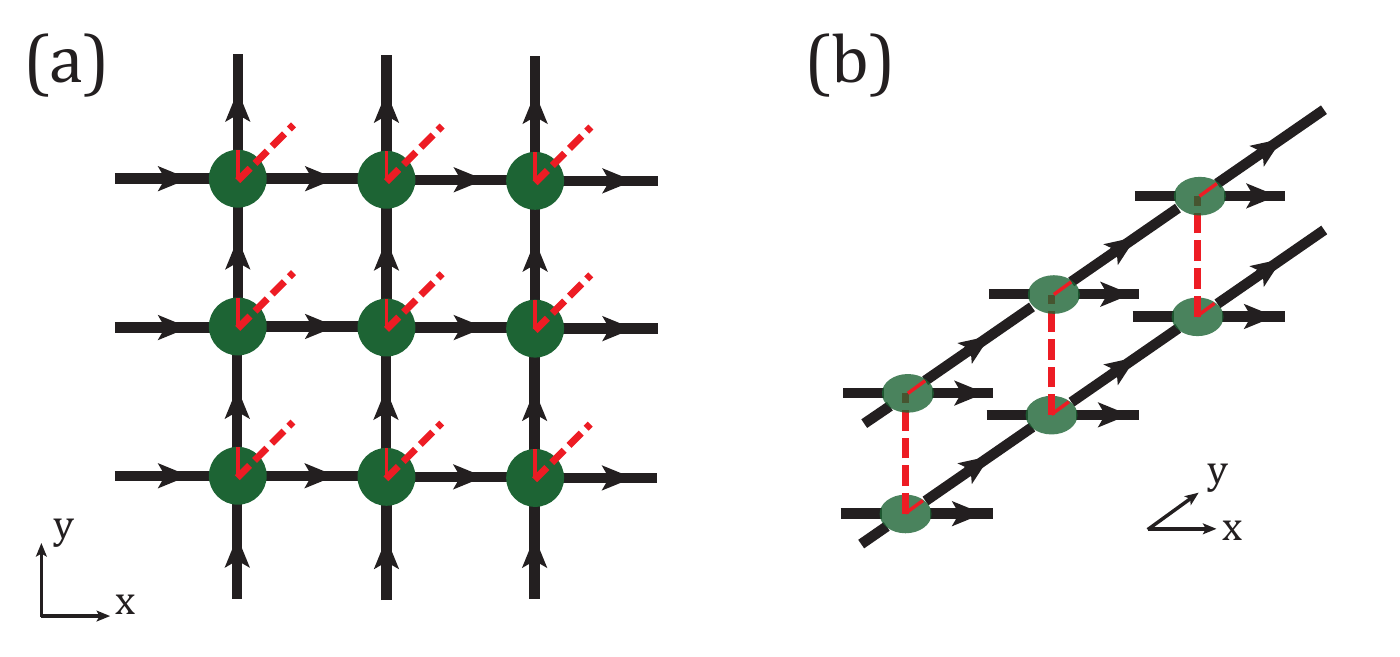}
}
\caption{\label{Square_PEPSnetwork1} (a) This graph represents the tensor network of the slave-spin PEPS $|\Psi_{D}\rangle$. There is a singlet projector $S_{\alpha\beta}$ on every link and tensor $D^m_{nesw}$ on every site. (b) shows the graphical representation of the horizontal transfer matrix (HTM). }
\end{figure}

Does the field theoretic obstacle strictly forbid the existence of a spin-1 featureless paramagnet? The answer is no: we can construct a slave-spin PEPS which serves as an example of a  spin-1 featureless paramagnet. We start with 4 auxiliary spin-1/2 degrees of freedom per site on the square lattice with each spin 1/2 attached to a different link connected to the same site. On each link, we project the two spin 1/2s into a singlet pair using $S_{\alpha\beta}$. The physical spin-1 function $|\Psi_{D}\rangle$ is obtained by a second projection that project the 4 auxiliary spin 1/2 degrees of freedom on each site to a total-spin-1 channel. In general, the projection from 4 spin-1/2 degrees of freedom to a total-spin-1 channel is not unique. However, there is only one choice that guarantees the $R_{\frac{2\pi}{4}}$, $\mathcal{T}$, $\sigma_\text{v}$ and $\sigma_{45^\circ }$ symmetries. Its graphical representation is shown in Fig. \ref{Square_SiteTensor1}, where we first form triplet pairs between the upper and lower spin 1/2s and between the left and right spin 1/2s and project the two triplet pairs into a total-spin-1 channel. The associated tensor is given by 
\begin{align}
D^m_{nesw} &=\sum_{n_{1,2}} \tilde{T}^m_{n_1n_2} T^{n_1}_{ew} T^{n_2}_{ns}.
\label{Square_Site}
\end{align}
where $n_{1,2}$ sum over $1$, $2$ and $3$ (which label the intermediate spin-1 states) and the tensor $T$ and $\tilde{T}$ are the CG coefficients previously defined.  The graphical representation of $|\Psi_{D}\rangle$ is now given by Fig. \ref{Square_PEPSnetwork1}.

$|\Psi_{D}\rangle$ is by construction translationally invariant. Since all of the tensors involved are the CG coefficients, the global $SO(3)$ spin rotation symmetry is also preserved. Under the site-centered or plaquette-centered $R_\frac{2\pi}{4}$ (say counter-clockwise), each tensor $D^m_{nesw}$ contributes a $-1$ phase factor because under the cyclic permutation of the subscripts
\begin{align}
D^m_{nesw} &= -D^m_{eswn}.
\label{Square_SiteRotate}
\end{align}
On top of that, the arrow on every horizontal link is flipped after the $\pi/2$ counter-clockwise rotation, each producing a $-1$ factor, while all the vertical ones are intact. Notice that the square lattice in a planar or torus geometry has equal number of sites and horizontal links. The state $|\Psi_{D}\rangle$ is invariant under $R_\frac{2\pi}{4}$ (with 0 lattice angular momentum). Since both $D^m_{nesw}$ and $S_{\alpha\beta}$ are real, $|\Psi_{D}\rangle$ is $\mathcal{T}$-invariant. We can also verify that $|\Psi_{D}\rangle$ preserves $\sigma_\text{v}$ and $\sigma_{45^\circ }$.

\begin{figure}
\centering
\begin{minipage}{.25\textwidth}
  \centering
  \includegraphics[width=1\linewidth]{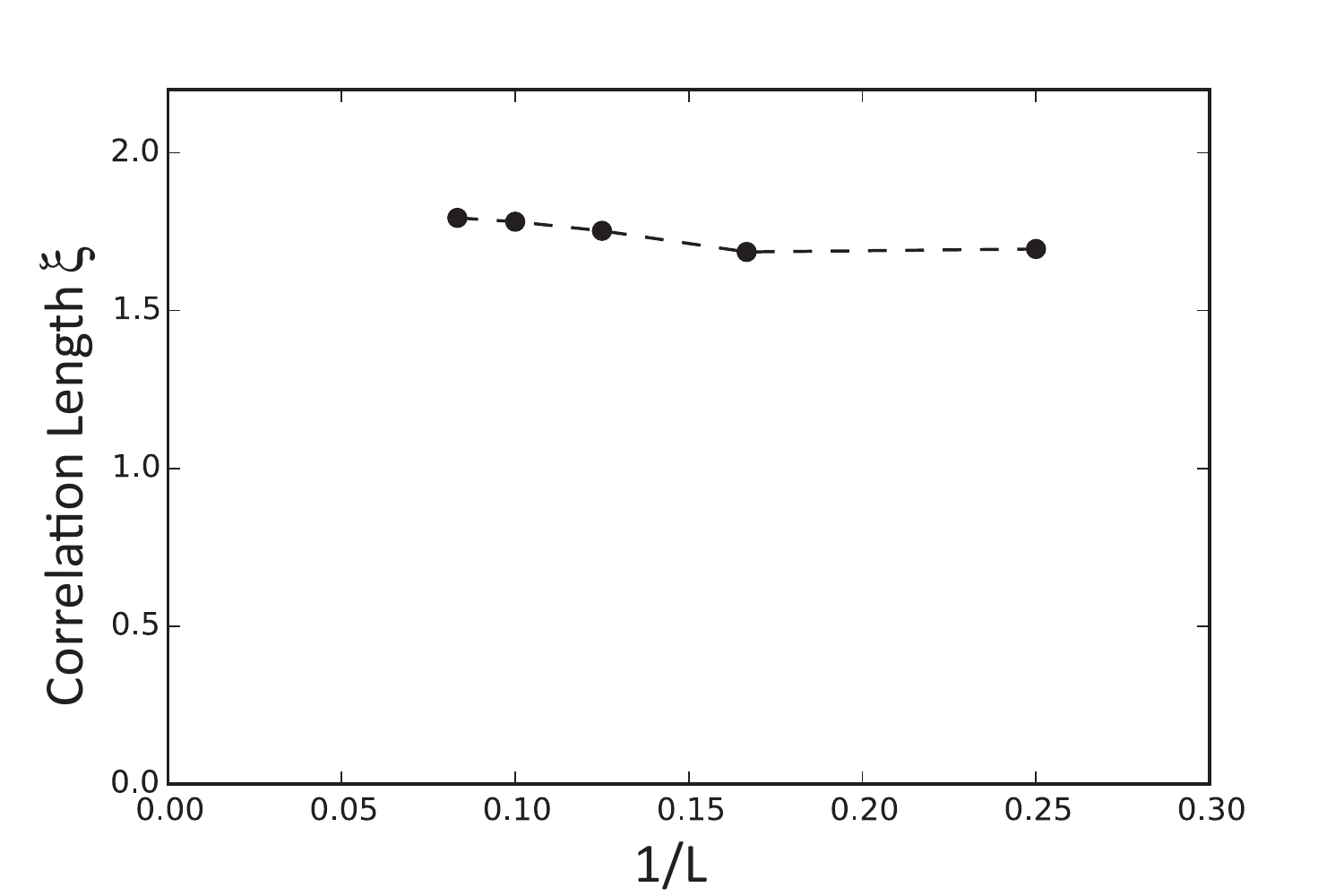}
\end{minipage}%
\begin{minipage}{.25\textwidth}
  \centering
  \includegraphics[width=1\linewidth]{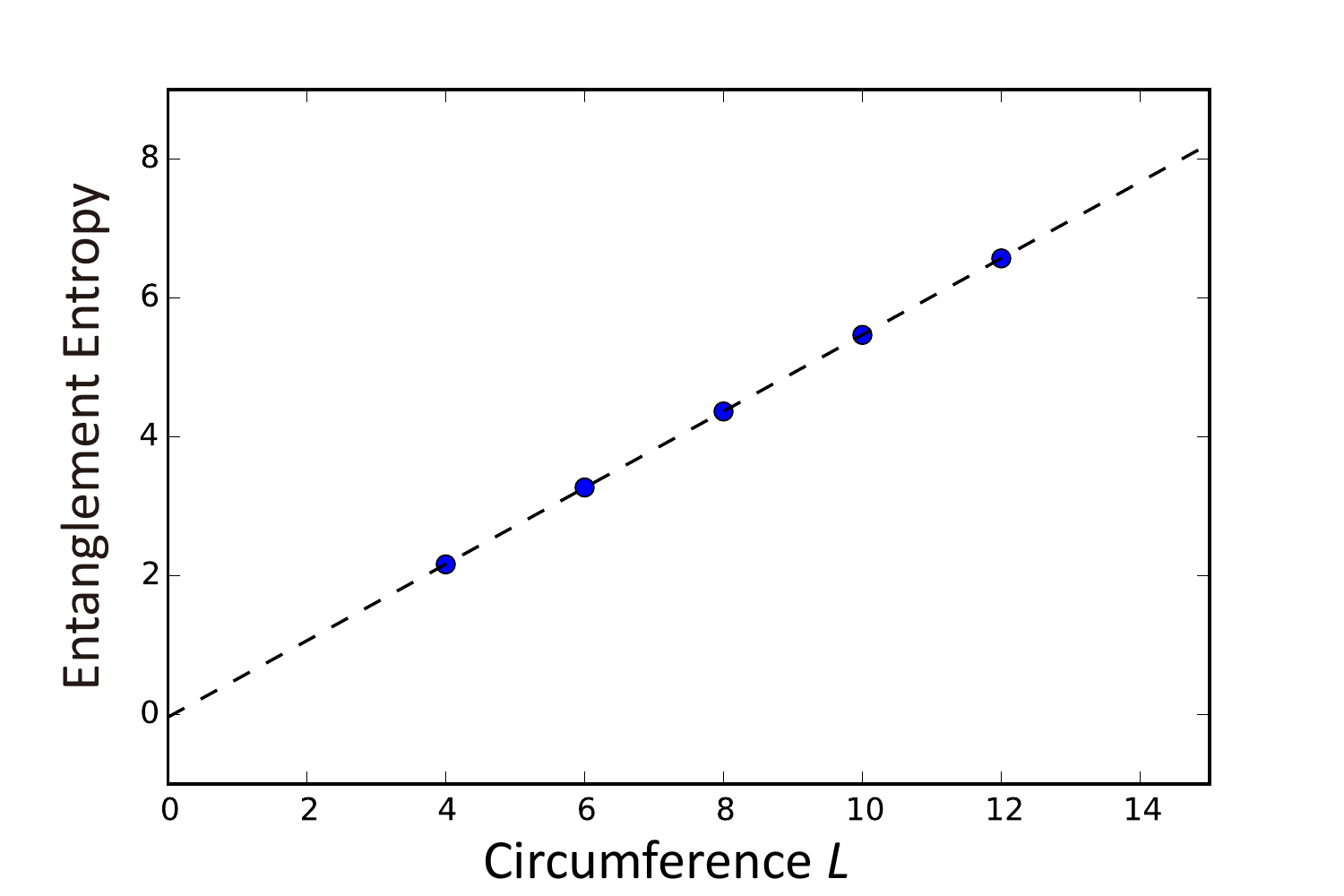}
  
\end{minipage}
\caption{Left panel: The physical correlation length $\xi$ in the unit of lattice spacing is plotted against $1/L$. The result is obtained from the DTM with $L=4,6,8,10,12$. The HTM result is similar. Right panel: Entanglement entropy of the semi-finite cylinder is plotted against the circumference $L$. A linear fit captures the ``area law" and the TEE.}
\label{Square_TRI_CorEE}
\end{figure}

At this point, we have shown that $|\Psi_{D}\rangle$ preserved all the required symmetries as a PEPS wave function. Now we verify numerically that (1) no symmetry is spontaneously broken (``cat state") and (2) no topological order is present in in $|\Psi_{D}\rangle$. We consider $|\Psi_{D}\rangle$ on a cylindrical geometry with finite circumference $L$ and view it as an MPS. As long as the transfer matrix of this MPS has a unique most dominant eigenvector (MDE) and the physical correlation length stays finite as $L\rightarrow \infty$, we can ensure the absence of spontaneously broken symmetries. Two types of cylindrical geometries with either the $y$ direction or the $45^\circ$ direction compact should be considered, which leads to two types of transfer matrices:  the horizontal transfer matrix (HTM) and the diagonal transfer matrix (DTM). We perform the exact diagonalization to the HTM and DTM for $L=2,3,...,12$. In each case, we obtain a unique MDE. The physical correlation length $\xi$ is plotted against $1/L$ in the left panel of Fig. \ref{Square_TRI_CorEE} and found to saturate as $1/L\rightarrow 0$. The uniqueness of the MDE for both transfer matrices and the finite correlation length ensure the absence of the spontaneous symmetry breaking for the ones compatible with either of the transfer matrices, which are the translation, the $SO(3)$ spin rotation, $\mathcal{T}$, $\sigma_\text{v}$ and $\sigma_{45^\circ }$. To verify the $R_\frac{2\pi}{4}$  symmetry, we calculated the nearest neighbor (NN) spin-spin correlation function along the 4 different directions. They agree with each other at the value $\langle\vec{S}\cdot\vec{S}\rangle_{NN}=-0.462 $ even though the cylindrical geometry is incompatible with $R_\frac{2\pi}{4}$. Therefore, we expect $R_\frac{2\pi}{4}$ is not spontaneously broken in the infinite-plane limit. All the symmetry properties of $|\Psi_{D}\rangle$ are now confirmed.

Moreover, the $R_\frac{2\pi}{4}$ symmetry of $|\Psi_{D}\rangle$ and the uniqueness of the MDE for both HTM and DTM also implies the absence of topological orders. The argument is the following. Let's first assume that $|\Psi_{D}\rangle$ is topologically ordered. We can construct the wave function $|\Psi_{D}\rangle$ on a $L_x \times L_y$ torus with periodic boundary conditions along along the $x$ and $y$ direction. When $L_x=L_y$, the $R_\frac{2\pi}{4}$ symmetry of $|\Psi_{D}\rangle$ implies that $|\Psi_{D}\rangle$ is an eigenstate of the topological $\mathsf{S}$ matrix \cite{ZhangF2012}. An eigenstate of the topological $\mathsf{S}$ matrix does not have a fixed anyon flux along either the $x$ or $y$ cycle of the torus. In other words, the state $|\Psi_{D}\rangle$ on the torus is not a minimally entangled state (MES)\cite{ZhangF2012} along either directions. Therefore, the Wilson loops around the $y$ ($x$) direction cycle develops a long-range order when we view the system as a quasi-1D system in the $x$ ($y$) direction. If the system is gapped, this long-range order will persist even when we open up the torus to create a infinite cylinder (with fixed circumference). The HTM will develop degeneracy in its MDEs. Similar argument can be made for the DTM. However, this is inconsistent with the numerical result which shows a unique MDE. Therefore, we prove by contradiction that the PEPS $|\Psi_{D}\rangle$ is not topologically ordered. To further justify this argument, we calculate the entanglement entropy (EE) of a semi-infinite cylinder and plot it against the circumference $L$ in Fig. \ref{Square_TRI_CorEE}. We obtain an ``area law" plus a topological entanglement entropy (TEE)  \cite{Kitaev2006EE,Levin2006EE} $\gamma \simeq 0$, which indicates the absence of topological order. Therefore, we conclude $|\Psi_{D}\rangle$ is indeed a featureless paramagnet.

\begin{figure}[tb]
\centerline{
\includegraphics[width=3.5
in]{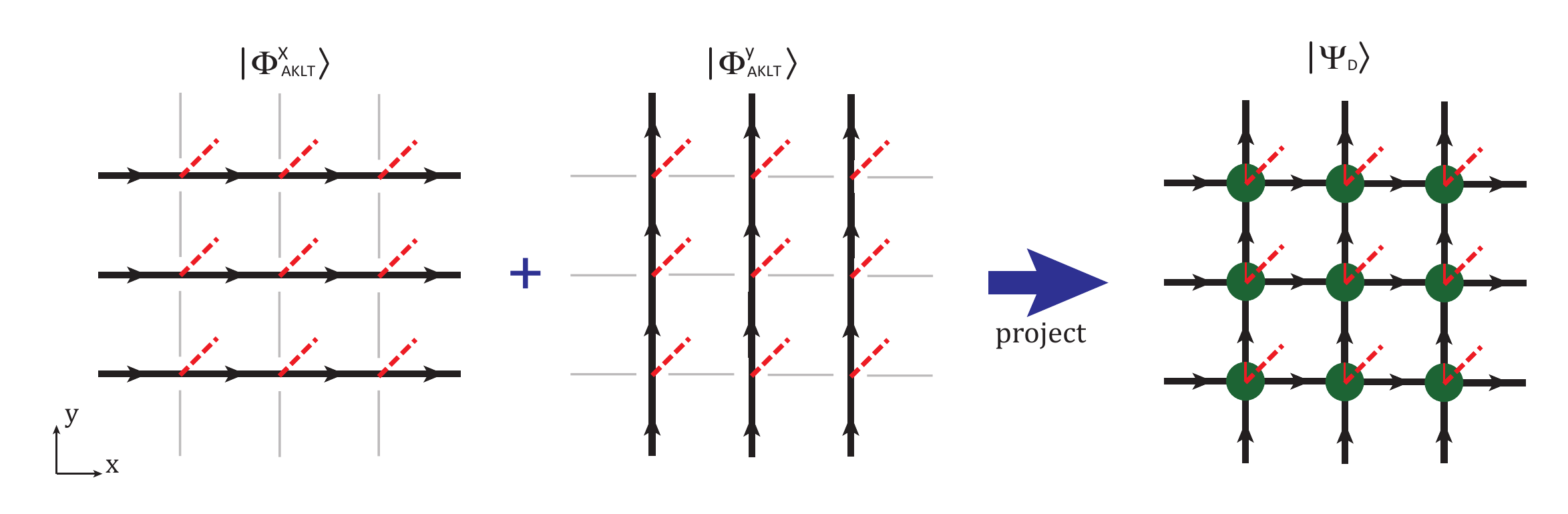}
}
\caption{\label{Square_Proj_Nematic} $|\Phi^x_\text{AKLT}\rangle$ ($|\Phi^y_\text{AKLT}\rangle$) is obtained by stacking 1D AKLT chain along the $y$ ($x$) direction. Both of them are nematic paramagnets without topological order. The slave-spin PEPS $|\Psi_{D}\rangle$ can be obtained by stacking the two states, $|\Phi^x_\text{AKLT}\rangle$ and $|\Phi^y_\text{AKLT}\rangle$, on top of each other and project the two spin 1s per site into a total-spin-1 channel.}
\end{figure}

Now, we have confirmed that $|\Psi_{D}\rangle$ is a featureless paramagnetic state which lies beyond the field theoretic expectation. Consequently, if there is a second-order transition separating the featureless paramagnet $|\Psi_{D}\rangle$ and the Neel order or the VBS order where the Berry phase pattern in Fig. \ref{Square_Phase_Pattern} is obtained, the critical theory should be a generalized version of the deconfined quantum criticality and most likely one that properly captures the ``slaved" degrees of freedom in the slave-spin PEPS. The construction of such a critical theory is beyond the scope of this paper and will be left for future works. Apart from the field theory, one can also try to search for simple Hamiltonians such that $|\Psi_{D}\rangle$ is an approximate ground state. In the numerics, we calculate the nearest neighbor (NN) and the next nearest neighbor (NNN) spin-spin correlation functions and obtain $\langle \vec{S}\cdot \vec{S}\rangle_\text{NN}  = -0.462$ and $ \langle \vec{S}\cdot \vec{S}\rangle_\text{NNN}  = -0.174$ 
%(see App. \ref{SSCF_Square} for details). 
Since both of them are anti-ferromagnetic (AF), a Hamiltonian with AF $J_1$ and $J_2$ Heisenberg couplings should be preferable. The actual $J_1$-$J_2$ Heisenberg model at intermediate $J_2/J_1$ is shown to host a nematic phase\cite{Jiang2009} that is captured by stacked AKLT chains\cite{Wang2015}. We denote the 2 symmetry related stacked-AKLT-chain states as $|\Phi^x_\text{AKLT}
\rangle$ and $|\Phi^y_\text{AKLT}\rangle$. Their tensor network representations are given in Fig. \ref{Square_Proj_Nematic}. When we stack $|\Phi^x_\text{AKLT}\rangle$ on top of $|\Phi^y_\text{AKLT}\rangle$ and project the two spin 1 degrees of freedom on each site in to a total-spin-1 channel, we can obtain the featureless paramagnet $|\Psi_{D}\rangle$ (see Fig. \ref{Square_Proj_Nematic}). It is an interesting question whether this connection between the wave functions leads to a dynamical theory that captures both the nematic phase and the featureless phase.

\section{2+1D Featureless Paramagnets on the Honeycomb Lattice} 
\label{Honeycomb}
In this section, we will study the featureless paramagnets for spin-1/2 and spin-1 honeycomb lattice with the following symmetries: the global spin rotation, the translation, the lattice $C_3$ rotation ($R_\frac{2\pi}{3}$), the time reversal ($\mathcal{T}$) and the 2 link-centered spatial mirror reflections symmetries ($\sigma_\bot$ with the mirror plan perpendicular and $\sigma_\|$ parallel to the link). We will first review the field theoretic obstacle to such featureless paramagnetic phases. Then, we prove the existence of the spin-1 featureless paramagnet by providing an example wave function in the slave-spin PEPS form. A numerical analysis verifies its featureless-ness. For the spin-1/2, we proposed two types of slave-spin PEPS with bond dimension $\chi=2,6$. The PEPSs with $\chi=2$ are found to be ``almost featureless paramagnets" with time reversal symmetry and at least one of the mirror reflection symmetries broken. Last but not least, we proposed the $\chi=6$ slave-spin PEPS for spin-1/2 honeycomb system. We show analytically that the slave-spin PEPS is fully symmetric. The numerical test for a gap is inconclusive, so we can only  conjecture that the proposed $\chi=6$ slave-spin PEPS is a featureless spin-1/2 paramagnet on the honeycomb lattice.

\subsection{Field Theoretic Obstacle to Paramagnetic Phases on the Honeycomb Lattice}
\begin{figure}[tb]
\centerline{
\includegraphics[width=3
in]{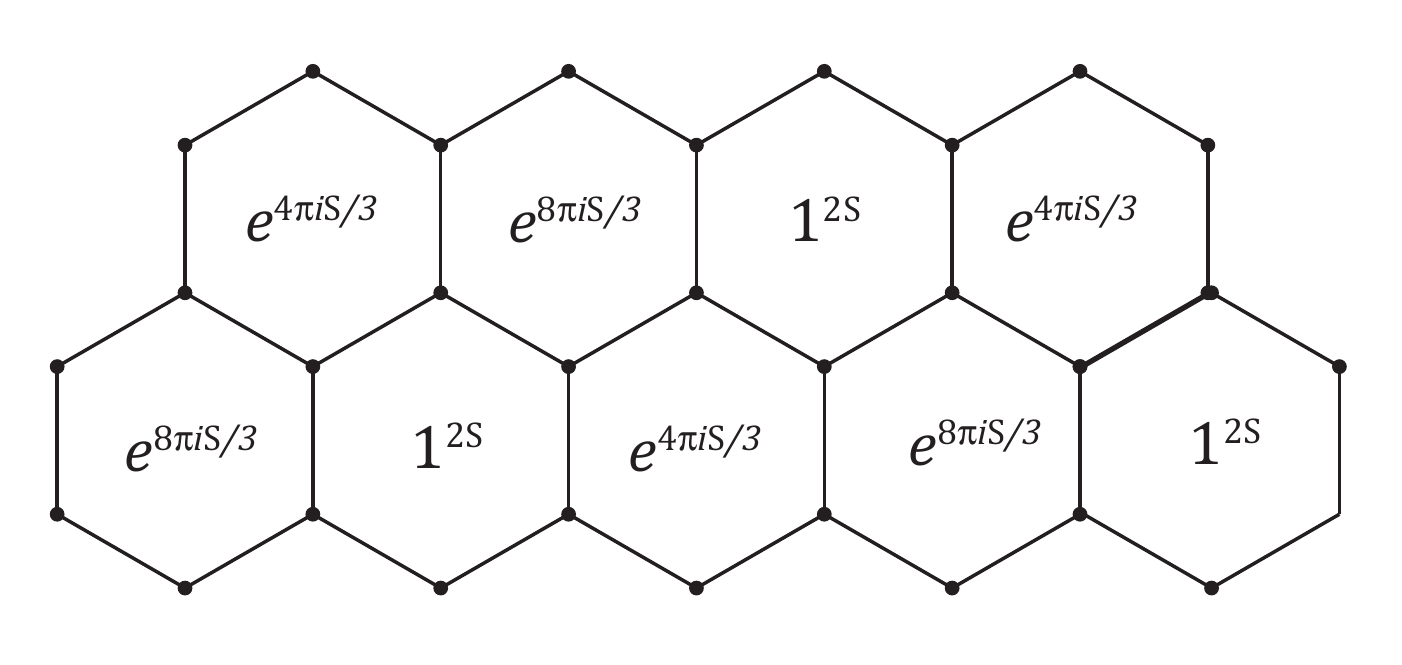}
}
\caption{\label{Honeycomb_Phase_Pattern} This plot shows the spatial dependence of the non-trivial monopole tunneling Berry phase in a spin-S Neel background on a honeycomb lattice. }
\end{figure}

Similar to strategy described in Sec. \ref{Square_Field_Theory}, we start with the classical Neel background with spin $\S$ per site and consider the proliferation of monopole tunneling events in the spacetime to obtain a paramagnetic phase. A generalized spin-coherent-state path integral or dimer model calculation \cite{Haldane1988,ReadSachdev1990,Zhao2012} shows that the monopole tunneling events on the honeycomb lattice have a non-trivial location dependent Berry phase pattern shown in Fig. \ref{Honeycomb_Phase_Pattern}. When the Neel background is disordered, this non-trivial Berry phase pattern rules out the possibility of spin-1 and spin-1/2 featureless paramagnets. In contrast, when $\S = \frac{3}{2} \mathbf{Z}$, the Berry phase pattern becomes trivial. Indeed, the most natural the AKLT state (a featureless paramagnet) on the honeycomb lattice requires $\S=3/2$\citep{AKLT1988}.

\subsection{Spin-1 Featureless Paramagnets on the Honeycomb Lattice}
\label{Honeycomb_Spin-1}
\begin{figure}[tb]
\centerline{
\includegraphics[width=3.5
in]{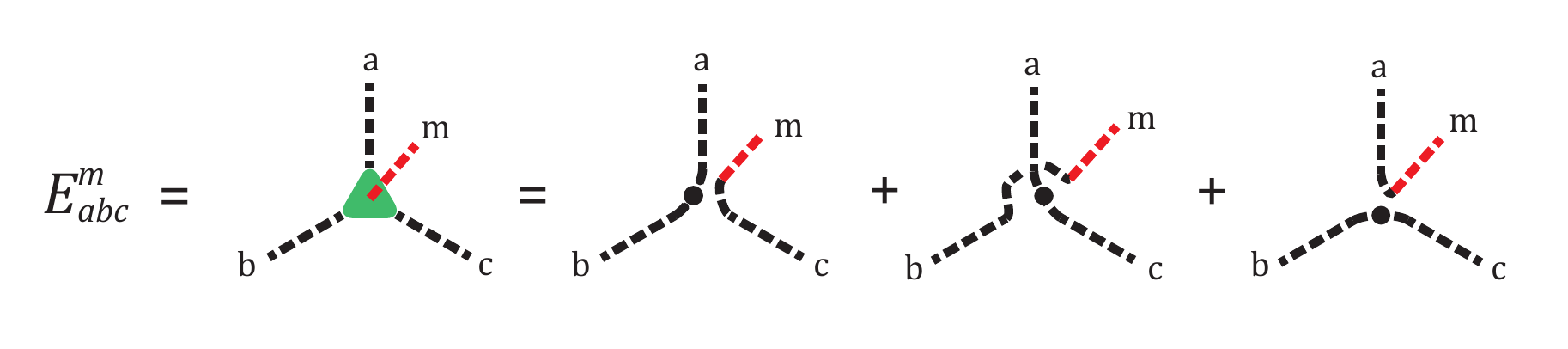}
}
\caption{\label{Honeycomb_Spin1_SiteTensor} 
The figure is the graphical representation of Eq. \ref{Honeycomb_Spin1_SiteTensorE}. The directly connected lines represent $\delta$-functions between the indices on their end points, while the lines decorated by a black dot represent the singlet projector $\tilde{S}$ defined previously. The physical degrees of freedom are represented by the red lines and auxiliary ones by the black lines.}
\end{figure}

\begin{figure}[tb]
\centerline{
\includegraphics[width=3.5
in]{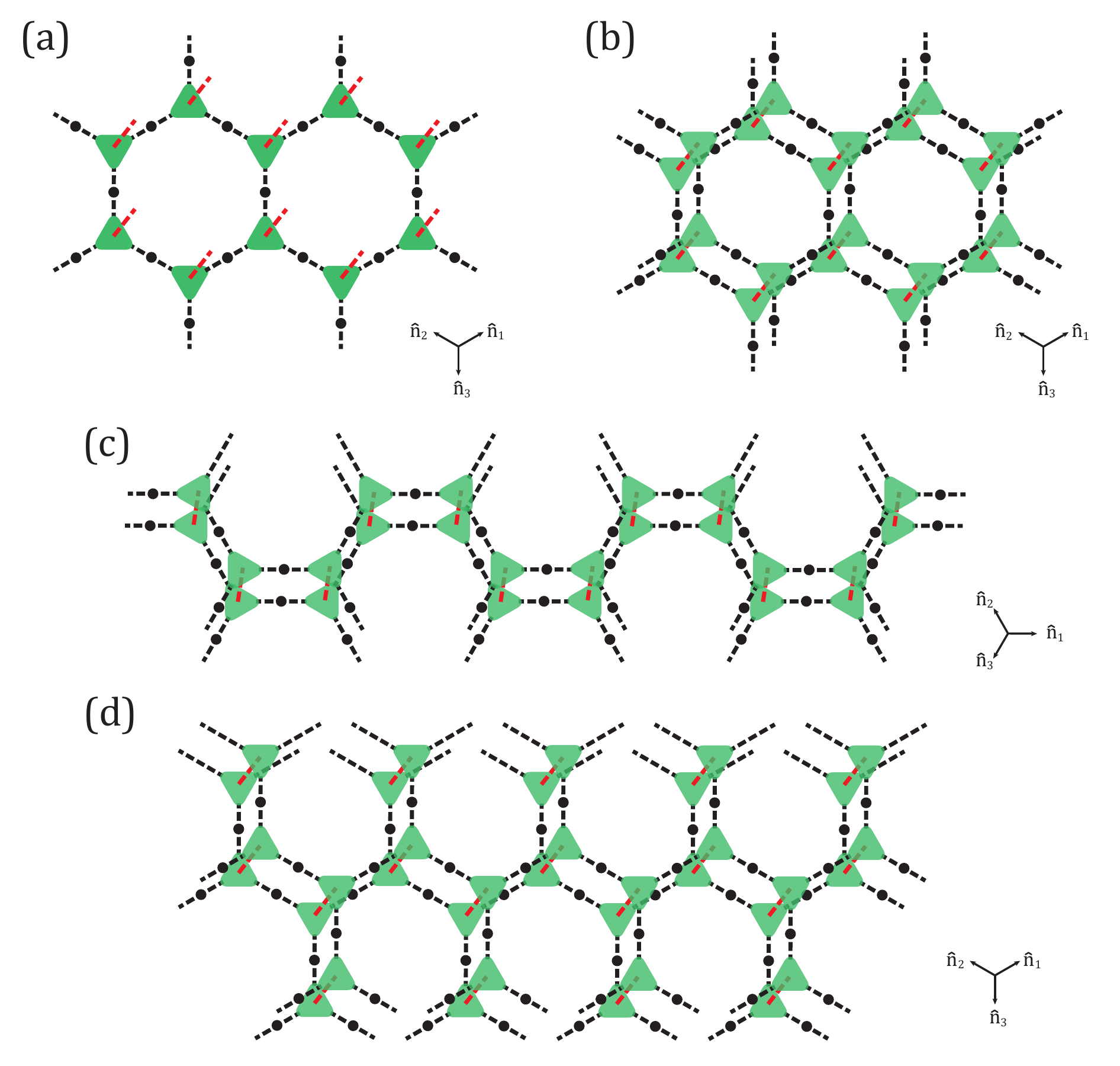}
}
\caption{\label{Honeycomb_Spin1_PEPSnetwork} 
(a) The tensor network for the wave function of $|\Psi^E_1\rangle$ is shown. $\vec{n}_{1,2,3}$ are unit vectors along the $30^\circ$, the $150^\circ$ and the $270^\circ$ directions. (b) The tensor network for the wave-function norm $\langle\Psi^E_1|\Psi^E_1\rangle$ is shown. When the physical degrees of freedom are first contracted, this tensor work can be viewed as a partition function of a classical statistical model on the auxiliary degrees of freedom. (c) The armchair transfer matrix (ATM) on an infinite plane is shown. (d) The zigzag transfer matrix (ZTM) on an infinite plane is shown.
}
\end{figure}

We prove the the existence of the spin-1 featureless paramagnet by writing down an example in the slave-spin PEPS form. We start with 3 auxiliary spin-1 degrees of freedom on each site with each spin-1 degree of freedom associated to each link that connects to the same site. Then we project the two spin-1 degrees of freedom associated to the same link into a total-spin-0 channel. This projection is implemented by the tensor $\tilde{S}_{nl}$ defined previously. Then, the three auxiliary spin-1 degrees of freedom on each site are projected onto a total-spin-1 channel to obtain a physical spin-1 wave function. The most symmetric choice of projection is given by the tensor
\begin{align}
E^{m}_{abc}= \tilde{S}_{ab} \delta_{mc}+\tilde{S}_{ac} \delta_{mb}+\tilde{S}_{bc} \delta_{mc},
\label{Honeycomb_Spin1_SiteTensorE}
\end{align}
whose graphical representation is given in Fig. \ref{Honeycomb_Spin1_SiteTensor}. Here, the indices $a$, $b$, $c$ and $m$ all run from 1 to 3 each representing 3 spin-1 states  in the $\hat{S}_z$ basis. The directly connected lines represent $\delta$ functions between the indices on their end points, while the lines decorated by a black dot represent the singlet projector $\tilde{S}$ defined previously. This tensor is invariant under the permutation of all the subscripts and $\mathcal{T}$:
\begin{align}
E^{m*}_{abc}=E^{m}_{abc}= E^{m}_{bca}= E^{m}_{acb}.
\label{Honeycomb_E_Permute}
\end{align}
After the second projection, we obtain the featureless paramagnetic wave function $|\Psi^E_1\rangle$, whose tensor network is shown in Fig. \ref{Honeycomb_Spin1_PEPSnetwork} (a). This tensor network has the tensor $\tilde{S}_{nl}$ on each link and the tensor $E^{m}_{abc}$ on each site.

\begin{figure}[tb]
\centerline{
\includegraphics[width=2.5
in]{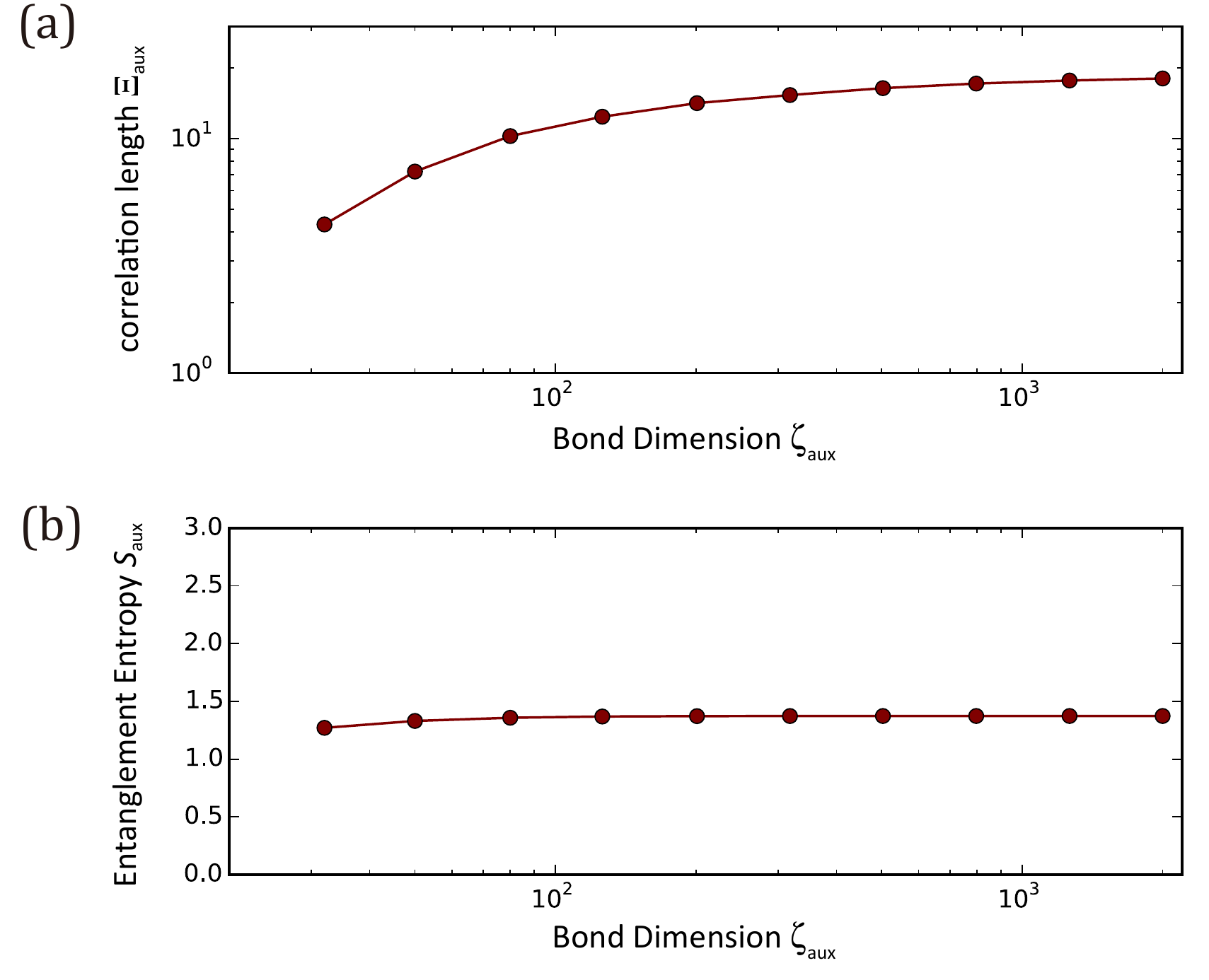}
}
\caption{\label{Honeycomb_Spin1_TEBD_atm} The correlation length $\Xi_\text{aux}$  of the auxiliary degrees of freedom (a) and the semi-infinite chain entanglement entropy $S_\text{aux}$ (b) of the MDE of the ATM are plotted against the bond dimension $\zeta_\text{aux}$. As we can see, both quantities have saturated at large $\zeta_\text{aux}$. The results for the ZTM are similar. 
}
\end{figure}

Because the tensors $E^{m}_{abc}$ and $\tilde{S}_{ab}$ are the CG coefficients that are real and invariant under the permutation of their subscripts, the wave function $|\Psi^E_1\rangle$ is invariant under all symmetries required. Now, we present the numerical evidence that $|\Psi^E_1\rangle$ is free of spontaneously broken symmetry and topological order. Due to its long correlation length, we cannot perform exact diagonalization study on finite-size cylinder. Instead, we directly study the planar limit. We view the wave-function norm $\langle\Psi^E_1|\Psi^E_1\rangle$ shown in Fig. \ref{Honeycomb_Spin1_PEPSnetwork} (b) as the partition function of a classical statistical model for the auxiliary spin-1 degrees of freedom.  Any spontaneous symmetry breaking manifests itself as the non-uniqueness of the MDE for its transfer matrices. For the honeycomb lattice, there are two types of transfer matrices: the armchair transfer matrix (ATM) (Fig. \ref{Honeycomb_Spin1_PEPSnetwork} (c)) and the zigzag transfer matrix (ZTM) (Fig. \ref{Honeycomb_Spin1_PEPSnetwork} (d)). The MDEs of the transfer matrices in the infinite-size limit is calculated using the infinite time-evolving block decimation (iTEBD) algorithm \cite{Vidal2007,Orus2008}. We denote the bond dimension in the iTEBD calculation as $\zeta_\text{aux}$. The correlation length $\Xi_\text{aux}$ of the the statistical model is plotted against $\zeta_\text{aux}$ in the upper panel of Fig. \ref{Honeycomb_Spin1_TEBD_atm}. When we view the MDE as a ground state of a 1D chain, we can measure semi-infinite chain entanglement entropy $S_\text{aux}$. Both $\Xi_\text{aux}$ and $S_\text{aux}$ saturate as $\zeta_\text{aux}$ increases, indicating the uniqueness of the MDE and finite-ness of the physical correlation length. Due to the geometry of the ATM and the ZTM, this result only suffices to prove that $|\Psi^E_1\rangle$ is invariant (without spontaneous symmetry breaking) under the translation with a doubled unit cell, the SO(3) spin rotation, $\mathcal{T}$, $\sigma_\bot$ and $\sigma_\|$. To show that $R_\frac{2\pi}{3}$ and fully translation symmetry is not spontaneously broken, we calculate the NN spin-spin correlation function along the $\hat{n}_1$, $\hat{n}_2$ and $\hat{n}_3$ (see Fig.\ref{Honeycomb_Spin1_PEPSnetwork}).  Their values agree with each other: $\langle \vec{S} \cdot \vec{S} \rangle_\text{NN} \simeq -0.110$. Therefore, the full translation symmetry and the $R_\frac{2\pi}{3}$ symmetry are not spontaneously broken. At this point, we have confirmed that the wave function $|\Psi^E_1\rangle$ preserve, without spontaneous symmetry breaking, all the required symmetries.

%The numerical simulation is performed up to $\zeta_\text{aux}=1200$. The correlation length $\Xi_\text{aux}$ of the auxiliary degrees of freedom in the the statistical model is shown in the upper panel of Fig. \ref{Honeycomb_Spin1_TEBD_atm}. When we view the transfer matrix as an imaginary time evolution operator and the MDE as the ground state of the 1D spin chain, we calculate the semi-infinite chain entanglement entropy $S_\text{aux}$. Both $\Xi_\text{aux}$ and $S_\text{aux}$ saturate as $\zeta_\text{aux}$ increases, indicating the uniqueness of the MDE and finite-ness of the physical correlation length. Due to the geometry of the ATM and the ZTM, this result only suffices to prove that $|\Psi^E_1\rangle$ is invariant (without spontaneous symmetry breaking) under the translation with a doubled unit cell, the SO(3) spin rotation, $\mathcal{T}$, $\sigma_\bot$ and $\sigma_\|$. To show that $R_\frac{2\pi}{3}$ and fully translation symmetry is not spontaneously broken, we calculate the NN spin-spin correlation function along the $\hat{n}_1$, $\hat{n}_2$ and $\hat{n}_3$. Their values agree with each other: $\langle \vec{S} \cdot \vec{S} \rangle_\text{NN} \simeq -0.110$. Therefore, the full translation symmetry and the $R_\frac{2\pi}{3}$ symmetry are not spontaneously broken. At this point, we have confirmed that the wave function $|\Psi^E_1\rangle$ preserve, without spontaneous symmetry breaking, all the symmetries including the (full) translation, the $R_\frac{2\pi}{3}$, the SO(3) spin rotation, $\mathcal{T}$, $\sigma_\bot$ and $\sigma_\|$.

With the uniqueness of the MDE of the transfer matrices and the $R_\frac{2\pi}{3}$ symmetry of $|\Psi^E_1\rangle$, we now argue that the state $|\Psi^E_1\rangle$ is free of any topological order. This argument is similar to the one provided in Sec. \ref{Square_FP}. We first assume that $|\Psi^E_1\rangle$ is topologically ordered. On a finite-size torus geometry compatible with $R_\frac{2\pi}{3}$, the invariance of $|\Psi^E_1\rangle$ under $R_\frac{2\pi}{3}$ implies that it is an eigenstate of $\mathsf{S}\mathsf{T}$, where $\mathsf{S}$ and $\mathsf{T}$ are the topological $\mathsf{S}$ matrix and $\mathsf{T}$ matrix. An eigenstate of $\mathsf{S}\mathsf{T}$ cannot be an MES along either non-trivial cycles of the torus. Consequently, the Wilson loops around one non-contractible cycle of the torus develops a long range order along the direction of the conjugate cycle. We argue that this long range order should persist even when the torus is deformed to a finite-circumference infinite cylinder, resulting in degeneracy of the MDEs of the cylinder transfer matrices. As we increase the circumference  $L$ and approach the planar limit, the degeneracy should not be lifted due to its topological nature. However, in the planar limit, the ATM and ZTM each only yields a unique MDE. Therefore, we prove by contradiction that $|\Psi^E_1\rangle$ is not topologically ordered and, thus, a spin-1 featureless paramagnet.

\subsection{Spin-1/2 Featureless Paramagnets on the Honeycomb Lattice}
In this subsection, we present the slave-spin PEPS study for the spin-1/2 featureless paramagnet on the honeycomb. This subsection contains two parts. In the first part, we focus on the slave-spin PEPS construction with bond dimension $\chi=2$. Due to the restriction of $\chi$, we only identify two "almost featureless paramagnets". $\mathcal{T}$ and at least one of $\sigma_\bot$ and $\sigma_\|$ is broken. Numerical studies verify their symmetry properties. In the second part, we propose a slave-spin PEPS construction for a spin-1/2 featureless paramagnet with bond dimension $\chi=6$.  

\subsubsection{Spin-1/2 slave-spin PEPS with $\chi=2$}
We start with 3 auxiliary spin-1/2 degrees of freedom on each site with each spin-1/2 degree of freedom associated to a different link that connects to the same site. Then we project the two spin-1/2 degrees of freedom associated to the same link into a singlet pair. Next, we project the 3 auxiliary spin-1/2 degrees of freedom on each site into a total-spin-1/2 channel to obtain a physical spin-1/2 wave function. There are 2 linearly independent total-spin-1/2 channels among 3 spin 1/2s, resulting in the 2 choices of tensors on the site:
\begin{align}
F^{\mu}_{\alpha\beta\gamma}= S_{\alpha\beta} \delta_{\mu\gamma} + e^{2\pi i/3}S_{\beta\gamma} \delta_{\mu\alpha} + e^{-2\pi i/3} S_{\gamma\alpha} \delta_{\mu\beta} , \nonumber \\
F'^{\mu}_{\alpha\beta\gamma}= S_{\alpha\beta} \delta_{\mu\gamma} + e^{-2\pi i/3}S_{\beta\gamma} \delta_{\mu\alpha} + e^{2\pi i/3} S_{\gamma\alpha} \delta_{\mu\beta}.
\label{Honeycomb_SpinH_Chi2_SiteTensorF}
\end{align}
Each of the indices $\mu$, $\alpha$, $\beta$ and $\gamma$ run from 1 to 2 representing the 2 states of a spin-1/2 degree of freedom. Under the cyclic permutation of the subscripts, they satisfy
\begin{align}
F^{\mu}_{\alpha\beta\gamma} & =e^{2\pi i/3} F^{\mu}_{\beta\gamma\alpha}, \nonumber \\
F'^{\mu}_{\alpha\beta\gamma} & =e^{-2\pi i/3} F'^{\mu}_{\beta\gamma\alpha}. \nonumber \\
\label{Honeycomb_SpinH_Chi2_SiteTensorRR}
\end{align}
Under $\mathcal{T}$ and ``mirror reflection", we have
\begin{align}
& F^{\mu}_{\beta\alpha\gamma}  = - F'^{\mu}_{\alpha\beta\gamma},  \nonumber \\
& (F^{\mu}_{\alpha\beta\gamma})^*  =  F'^{\mu}_{\alpha\beta\gamma}.
\label{Honeycomb_SpinH_Chi2_TMFF}
\end{align} 
To construct the slave-spin PEPS on the honeycomb lattice, we can either use the same tensor, say $F^{\mu}_{\beta\alpha\gamma}$, on every site or use two different tensors for the 2 sub-lattice of the honeycomb lattice. We denote the PEPS associated to the first option as $|\Psi^{FF}_{1/2}\rangle$ and the second option as $|\Psi^{FF'}_{1/2}\rangle$. We can check using Eq. \ref{Honeycomb_SpinH_Chi2_SiteTensorRR} and \ref{Honeycomb_SpinH_Chi2_TMFF} that the PEPS $|\Psi^{FF}_{1/2}\rangle$ and $|\Psi^{FF'}_{1/2}\rangle$ preserve the translation, the global spin rotation and $R_\frac{2\pi}{3}$. Regarding $\mathcal{T}$, $\sigma_\bot$ and $\sigma_\|$, $|\Psi^{FF}_{1/2}\rangle$ only preserves the combination of any two.  $|\Psi^{FF'}_{1/2}\rangle$ is only invariant under $\mathcal{T}\sigma_\|$ and $\sigma_\bot$. Therefore, we refer to them as almost featureless paramagnets.

\begin{figure}[tb]
\centerline{
\includegraphics[width=3in]{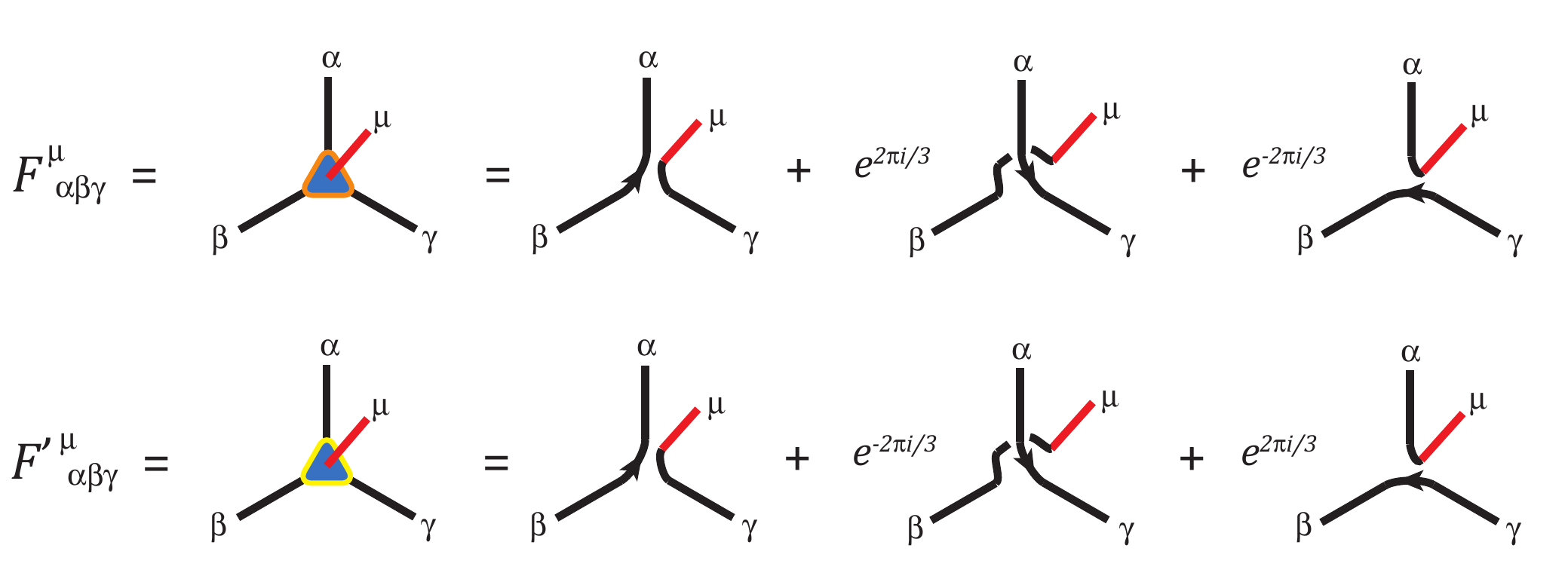}
}
\caption{\label{Honeycomb_SpinH_Chi2_SiteTensor} 
This figure is the graphical representation of Eq.\ref{Honeycomb_SpinH_Chi2_SiteTensorF}. The directly connected lines represent $\delta$-functions between the indices on their end points, while the lines decorated by an arrow represent the singlet projector $S$ defined in Eq. \ref{CG_HHto0}. The physical 1/2 degrees of freedom are represented by the red lines and auxiliary ones by the black lines.}
\end{figure}

\begin{figure}[tb]
\centerline{
\includegraphics[width=1.5
in]{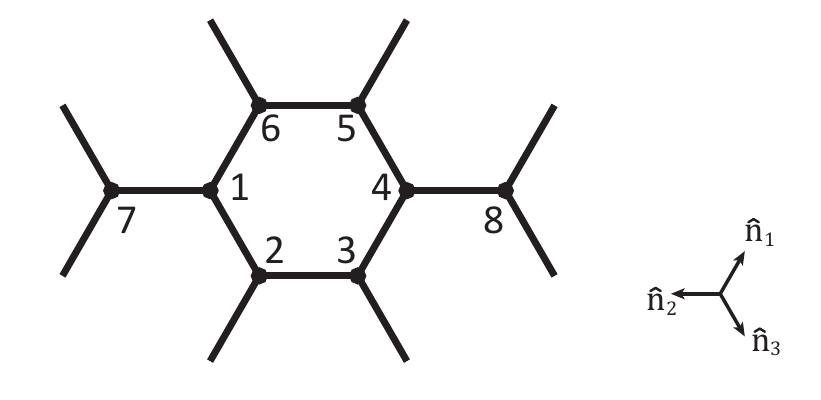}
}
\caption{\label{Honeycomb_SpinH_Chi2_sch} Different correlation functions including spin-spin correlation functions and the triple spin chiralities are calculated to verify the unbroken symmetries. Here, $\hat{n}_{1,2,3}$ are the unit vectors along the corresponding directions. 
}
\end{figure}

\begin{figure}
\centering
\begin{minipage}{.25\textwidth}
  \centering
  \includegraphics[width=1\linewidth]{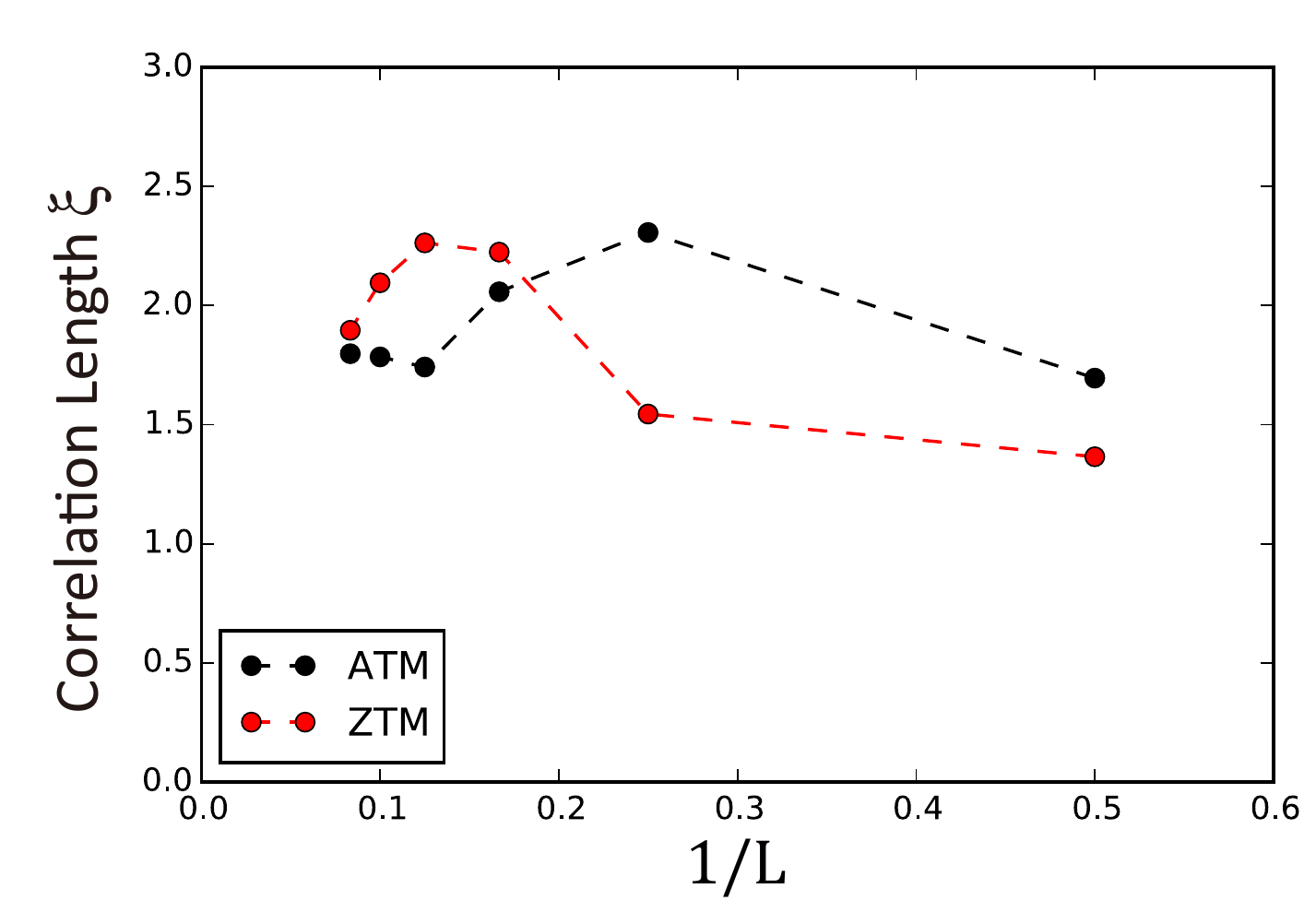}
\end{minipage}%
\begin{minipage}{.25\textwidth}
  \centering
  \includegraphics[width=1\linewidth]{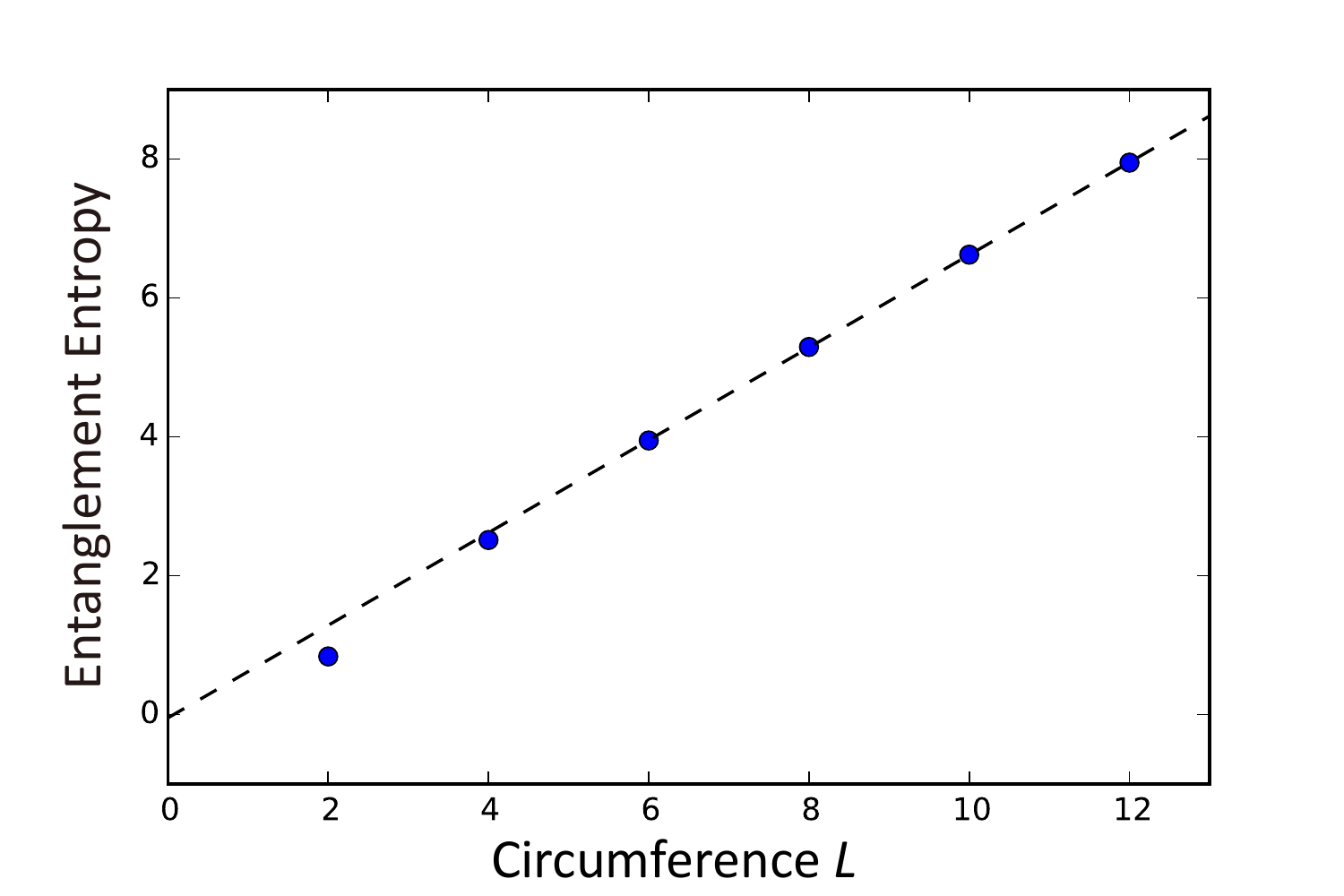}

\end{minipage}
\caption{Left panel: The physical correlation lengths $\xi$ (in the unit of lattice spacing) obtained from the ATM (black) and the ZTM (red) are plotted against $1/L$. Right panel: The semi-infinite cylinder entanglement entropy is plotted against the circumference $L$. The dash line is a linear fit that extracts the ``area law" and the TEE $\gamma$.}
\label{Honeycomb_SpinH_Chi2_aa}
\end{figure}

{\it Numerical study for $|\Psi^{FF}_{1/2}\rangle$ - } 
Here, we show numerically that $|\Psi^{FF}_{1/2}\rangle$, being an almost featureless paramagnet, does not suffer from spontaneous symmetry breaking and topological order. With $\chi=2$, we are able to perform diagonalization of the ATM and ZTM on finite circumference cylinder. For the circumference $L=2,4,6,8,10,12$, both the ATM and the ZTM has a unique MDE in each case. The physical correlation length $\xi$ is plotted against $1/L$ in left panel of Fig. \ref{Honeycomb_SpinH_Chi2_aa}. $\xi$ saturates to a finite value as $1/L\rightarrow0$ . Since the transfer matrices are not compatible with certain symmetries, this result only suffices to rule out the spontaneous breaking of the translation with a doubled unit cell and the global spin rotation symmetry. To verify the full translation and the $R_\frac{2\pi}{3}$ symmetry, we calculated the 3 NN spin-spin correlation functions along three different directions $\hat{n}_1$, $\hat{n}_2$, and $\hat{n}_3$. Their value agree with each other: $\langle \vec{S} \cdot \vec{S} \rangle_\text{NN} \simeq -1.08$. Therefore, the full translation symmetry and the $R_\frac{2\pi}{3}$ are confirmed. For the symmetries $\mathcal{T}\sigma_\bot$, $\mathcal{T}\sigma_\|$ and $\sigma_\|\sigma_\bot$, we calculated the symmetry related triple spin chiralities: $C_{ijk}\equiv\langle\vec{S}(\vec{r}_i) \cdot(\vec{S}(\vec{r}_j) \times\vec{S}(\vec{r}_k))\rangle$(see Fig. \ref{Honeycomb_SpinH_Chi2_sch} for the reference the subscripts) and find the following agreement:
\begin{align}
& C_{561} \xleftrightarrow{\sigma_\|\sigma_\bot} C_{716}  \xleftrightarrow{\mathcal{T}\sigma_\|}  -C_{712} \xleftrightarrow{\mathcal{T}\sigma_\bot} C_{561}, 
\nonumber \\
& C_{561}  = C_{716}  = -C_{712} =-0.05970 \pm 0.00002.
\end{align}
With this agreement plus other checks above, we verify that $|\Psi^{FF}_{1/2}\rangle$ indeed preserves (without spantenous symmetry breaking) the following symmetry: the global spin rotation symmetry, the translation symmetry, the $R_\frac{2\pi}{3}$ symmetry, $\sigma_\|\sigma_\bot$, $\mathcal{T}\sigma_\bot$ and $\mathcal{T}\sigma_\|$. 

With the uniqueness of the MDE of the transfer matrices and the invariance under $R_\frac{2\pi}{3}$, we can argue, following similar strategy in Sec. \ref{Honeycomb_Spin-1} and in Sec. \ref{Square_FP}, that $|\Psi^{FF}_{1/2}\rangle$ is free of topological order. Nevertheless, we also calculate the entanglement entropy of the semi-infinite cylinder with circumference $L$ (see the right panel of Fig. \ref{Honeycomb_SpinH_Chi2_aa}) from which we obtain an ``area law" and the TEE $\gamma\sim 0$. Therefore, we confirm again the absence of topological order in $|\Psi^{FF}_{1/2}\rangle$. In summary, $|\Psi^{FF}_{1/2}\rangle$ is indeed an almost featureless spin-1/2 paramagnet.

\begin{figure}
\centering
\begin{minipage}{.25\textwidth}
  \centering
  \includegraphics[width=1\linewidth]{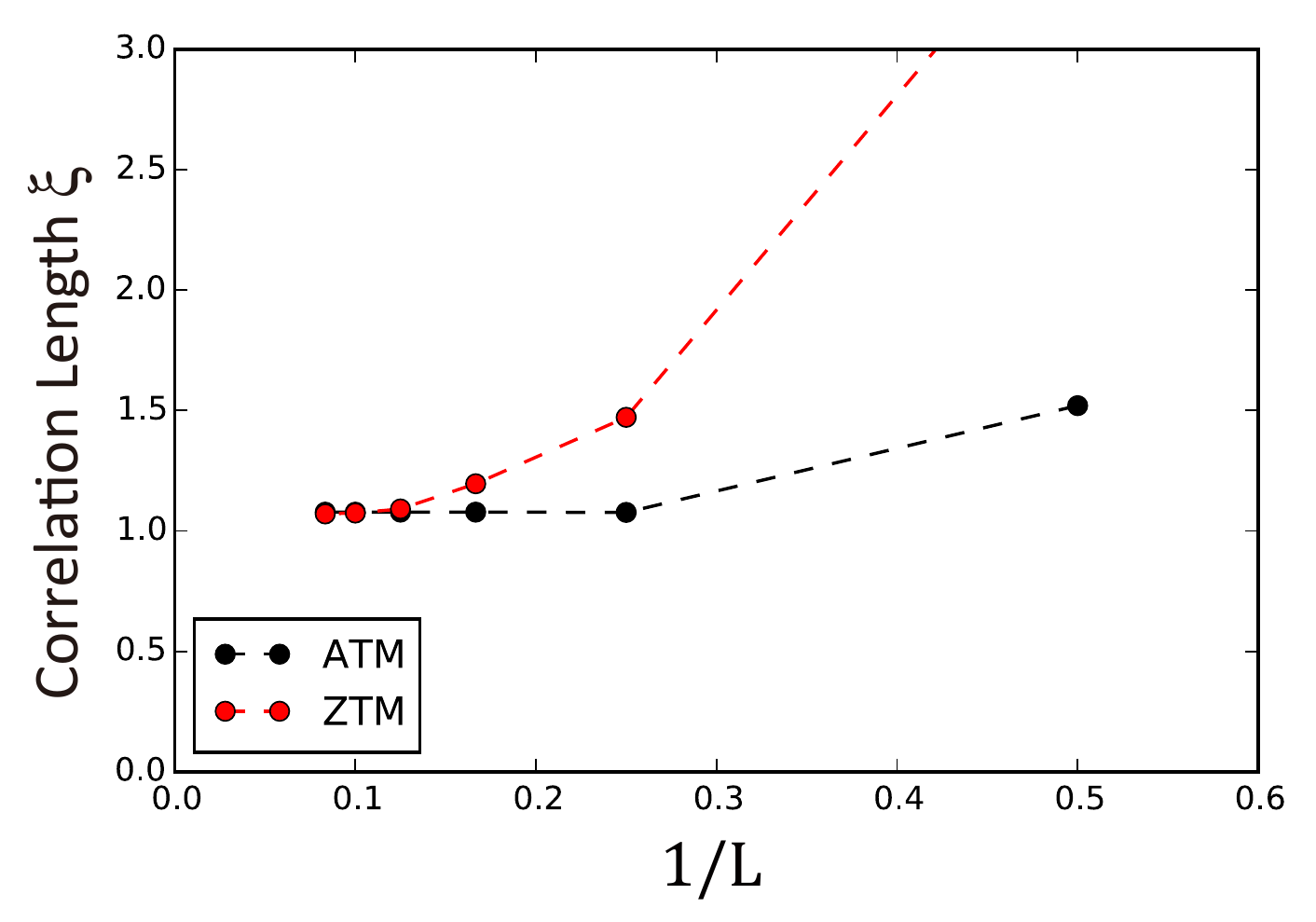}
\end{minipage}%
\begin{minipage}{.25\textwidth}
  \centering
  \includegraphics[width=1\linewidth]{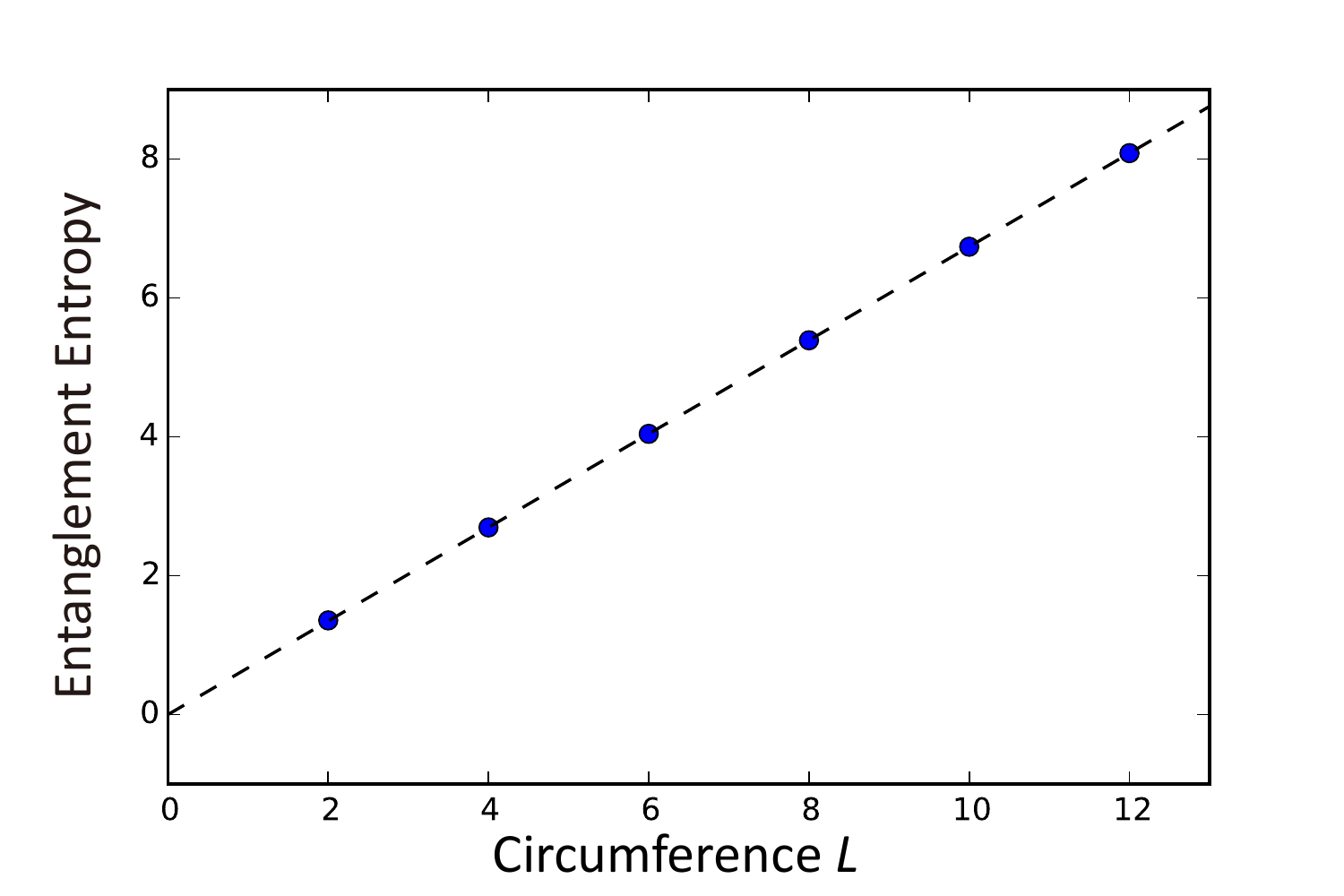}

\end{minipage}
\caption{Left panel: The physical correlation length $\xi$ (in the unit of lattice spacing) obtained from the ATM (black) and the ZTM (red) are plotted against $1/L$. Right panel: The semi-infinite cylinder entanglement entropy is plotted against the circumference $L$. The dash line is a linear fit that extracts the ``area law" and the TEE $\gamma$.}
\label{Honeycomb_SpinH_Chi2_ab}
\end{figure}

{\it Numerical study for $|\Psi^{FF'}_{1/2}\rangle$ - } 
Here we present the numerical study for $|\Psi^{FF'}_{1/2}\rangle$ on the cylinder with circumference $L=2,4,6,8,10,12$. Exact diagonalization of the ATM and ZTM yields a unique MDE in each case. The physical correlation length $\xi$ is plotted against $1/L$ in the left panel of Fig. \ref{Honeycomb_SpinH_Chi2_ab}. $\xi$ saturates to a finite value as $1/L\rightarrow0$. With these results, we establish the absence of the spontaneous breaking of the translation with a doubled unit cell, the global spin rotation symmetry and $\sigma_\bot$ symmetry. We also calculated the 3 NN spin-spin correlation functions along three different directions $\hat{n}_1$, $\hat{n}_2$, and $\hat{n}_3$. Their value agree with each other: $\langle \vec{S} \cdot \vec{S} \rangle_\text{NN} \simeq -0.076$. Therefore, the full translation symmetry and the $R_\frac{2\pi}{3}$ symmetry are not spontaneously broken. Regarding the $\mathcal{T}\sigma_\|$ symmetry, we calculate the symmetry related triple spin chiralities:
\begin{align}
& C_{716} \xleftrightarrow{\mathcal{T}\sigma_\|} - C_{712},  \nonumber \\
C_{716}  =& -C_{712}  =-0.0493 \pm 0.0003.
\end{align}
With this agreement plus other checks above, we verify that $|\Psi^{FF'}_{1/2}\rangle$ indeed preserves (without spontaneous symmetry breaking) the following symmetry: the global $SO(3)$ spin rotation symmetry, the translation symmetry, the $R_\frac{2\pi}{3}$ symmetry, $\sigma_\bot$, and $\mathcal{T}\sigma_\|$. 

Similarly, the uniqueness of the MDE of the transfer matrices and the invariance under $R_\frac{2\pi}{3}$ implies the absence of topological order in $|\Psi^{FF}_{1/2}\rangle$. Nevertheless, we calculate the entanglement entropy of the semi-infinite cylinder with finite circumference $L$ (see the right panel of Fig. \ref{Honeycomb_SpinH_Chi2_ab}) from which we obtain an ``area law" and the TEE $\gamma\sim 0$. Therefore, we confirm again the absence of topological order in $|\Psi^{FF'}_{1/2}\rangle$. In summary, $|\Psi^{FF'}_{1/2}\rangle$ is indeed an almost featureless spin-1/2 paramagnet.

To sum up this subsection, the two slave-spin PEPSs $|\Psi^{FF}_{1/2}\rangle$ and $|\Psi^{FF'}_{1/2}\rangle$ are both almost featureless paramagnets on the honeycomb lattice with spin 1/2 per site (with different symmetries). Before we end this section, we briefly comment on the possible Hamiltonians for these two states or for the phases they represent. As is mentioned above, both $|\Psi^{FF}_{1/2}\rangle$ and $|\Psi^{FF'}_{1/2}\rangle$ have AF NN spin-spin correlation functions. The NNN spin correlation functions are also anti-ferromagnetic: $\langle \vec{S}_1\vec{S}_5\rangle^{FF}_\text{NNN}\simeq -0.035$ and 
$\langle \vec{S}_1\vec{S}_5\rangle^{FF'}_\text{NNN}\simeq -0.041$. Therefore, these states would be favored by Hamiltonians with AF $J_1$ and $J_2$ Heisenberg couplings.

\subsubsection{Spin-1/2 slave-spin PEPS with $\chi=6$}
In the previous subsection, we study the slave-spin PEPSs with the bond dimension $\chi=2$ and find two almost featureless paramagnets. In this section, we will propose a slave-spin PEPS with $\chi=6$ for a fully featureless spin-1/2 paramagnet for the honeycomb lattice. We start with an auxiliary spin-1 and an auxiliary spin-3/2 degrees of freedom on each site of the honeycomb lattice. We put the spin-1 degrees of freedom on each site in the wave functions $|\Psi^E_1\rangle$ studied in Sec. \ref{Honeycomb_Spin-1}. Then we construct the AKLT state on the honeycomb lattice \citep{AKLT1988, Wei2011} using the the spin-3/2 degrees of freedom on each site. Both of $|\Psi^E_1\rangle$ and the AKLT state are featureless. The physical spin-1/2 wave function (denoted as $|\Psi_{1/2}\rangle$) is obtained by projecting the spin-1 and the spin-3/2 degrees of freedom on the same site into a total-spin-1/2 channel. Since $|\Psi^E_1\rangle$  and the spin-3/2 AKLT  are both PEPS with bond dimension 3 and 2, $|\Psi_{1/2}\rangle$ is a PEPS with $\chi=6$. $|\Psi_{1/2}\rangle$ is, by construction, invariant under the global $SO(3)$ spin rotation, the translation, the $R_\frac{2\pi}{3}$, $\mathcal{T}$, $\sigma_\bot$ and $\sigma_\|$.

\begin{figure}[tb]
\centerline{
\includegraphics[width=2.5
in]{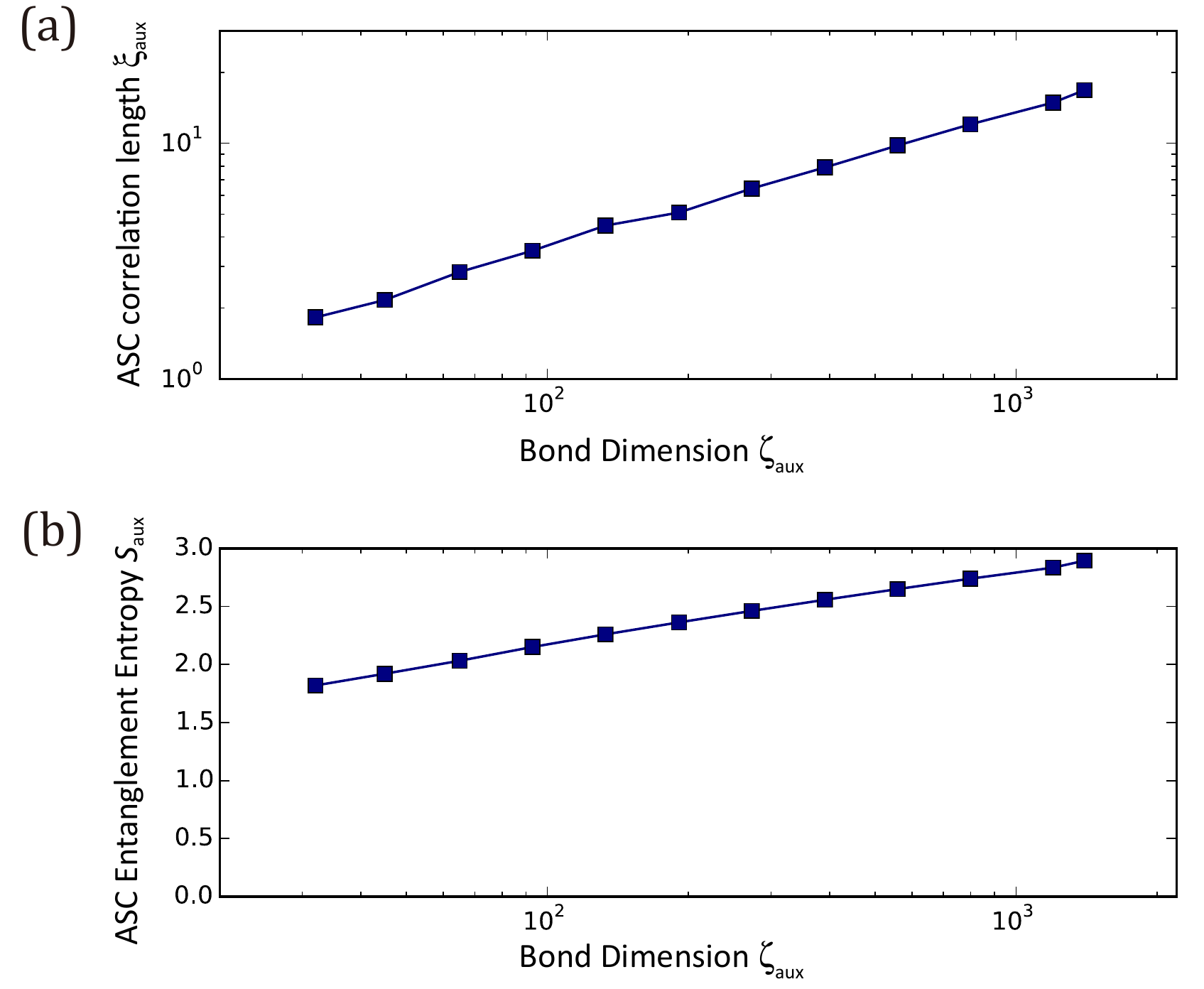}
}
\caption{\label{Honeycomb_SpinH_TEBD_atm} The correlation length $\Xi_\text{aux}$  of the auxiliary degrees of freedom (a) and the semi-infinite chain entanglement entropy $S_\text{aux}$ (b) of the MDE of the ATM are plotted against the bond dimension $\zeta_\text{aux}$. As we can see, both quantities haven't saturated even at the largest bond dimension $\zeta_\text{aux}=1200$.
}
\end{figure}

We perform the iTEBD study for $|\Psi_{1/2}\rangle$. The result is shown in Fig. \ref{Honeycomb_SpinH_TEBD_atm}. As we increase the bond dimension $\zeta_\text{aux}$ in the simulation, both the correlation length $\Xi_\text{aux}$ for the auxiliary degrees of freedom and the entanglement entropy $S_\text{aux}$ of the effective 1D chain given by the MDE grows linearly. With this result, we cannot distinguish the two possibilities: (1) the transfer matrix has a gapless spectrum or (2) the spectrum is gapped but the MDE need to be captured by  $\zeta>1200$. This leaves an interesting question open: whether or not $|\Psi_{1/2}\rangle$ is a featureless paramagnet. Note that Ref. \onlinecite{Kimchi2013} proposed a featureless wave function $|\Psi_{e\sigma}\rangle$ of spin-1/2 fermions at site filling 1 on the honeycomb lattice. Naively one could Gutzwiller project this wave function to produce a featureless spin-1/2 wave function, but they find the Gutzwiller projection actually annihilates the wave function: a spin-1/2 paramagnet is surprisingly evasive. In fact, the fermionic wave function $|\Psi_{e\sigma}\rangle$ is not even adiabatically connected to our proposal $|\Psi_{1/2}\rangle$ while preserving the lattice symmetries. That is because the two wave functions have opposite eigenvalues under $\sigma_\bot$ on an $L \times L$ torus with $L$ odd. Hence, $|\Psi_{e\sigma}\rangle$ and $|\Psi_{1/2}\rangle$ belong to different classes of fragile Mott insulators\cite{YaoKivelson} that are separated from each other by a phase transition mandated by the lattice symmetries.

\section{Conclusion and Discussion}
\label{Conclusion}
In this paper, we study the problem of constructing featureless paramagnets on the spin-1 square lattice and the spin-1 or spin-1/2 honeycomb lattice in 2+1D. The featureless paramagnets we focus on are short-range gapped states with certain symmetries including the global $SO(3)$ spin symmetry, the translation symmetry, the lattice rotation symmetry, the time reversal symmetry and other possible point group symmetries. Even though these featureless paramagnets are not forbidden by the LSM theorem or its generalizations, there are strong field theoretic obstacles to construct them. Nevertheless, to prove their existence, we use the slave-spin PEPS construction to write down ideal wave functions for the featureless paramagnets on the square lattice and honeycomb lattice. For the spin-1 slave-spin PEPS wave functions on the square lattice and the honeycomb lattice, we provide both analytical and numerical studies to verify their featureless-ness. For the honeycomb lattice with spin 1/2 per site, we present two types of construction of slave-spin PEPS with bond dimension $\chi=2$ and $\chi=6$. With $\chi=2$, we propose and verify two different almost featureless paramagnets with broken time reversal symmetry and the mirror reflection symmetries. With $\chi=6$, we propose a completely featureless slave-spin PEPS wave function. However, due to the lack to numerical evidence, the status of the spin-1/2 honeycomb model remains ambiguous.

All these featureless paramagnets are given in PEPS forms. Parent Hamiltonians can be constructed such that they are the exact ground states\cite{Perez-Garcia2008}. Sometimes the parent Hamiltonians take very complicated forms. It would be interesting to find simple Hamiltonians such that their ground states are well-approximated by or belong the same phase as these ideal featureless paramagnetic wave functions. We've provided hints to what terms to include in the Hamiltonians so that these ideal wave functions are favorable. As we've shown, for the the case of the spin-1 featureless paramagnet on the square lattice and the spin-1/2 almost featureless paramagnets on the honeycomb lattice, the NN and NNN spin-spin correlation functions are shown to be anti-ferromagnetic. Therefore, the Hamiltonians with AF $J_1$ and $J_2$ Heisenberg couplings is a good direction to look at.

As we've pointed out, the Berry phase pattern\cite{Haldane1988,ReadSachdev1990} in the spin-coherent-state path integral provides obstacles to the existence of these featureless paramagnets studied in the paper. In fact, the Berry phase pattern are obtained in the vicinity of the Neel order\cite{Haldane1988} and the valence-bond-solid \cite{ReadSachdev1990}. One interesting question would be whether it is possible to have a direct second order phase transition from these featureless paramagnetic phases to the Neel order or the VBS order. More generally, one can ask what are the classical orders that is separated from the featureless paramagnetic phases by a second order transition and how one should describe the transition in the field theoretic language. In this paper, we obtain the ideal wave functions from the slave-spin PEPS construction. In the vicinity of a second transition, if the slaved-spin degrees of freedom becomes dynamical, one might need a field theory description that is beyond both the Landau paradigm and that is a generalized version of the deconfined quantum criticality.

{\it Acknowledgement - } CMJ is indebted to the inspiring conversations with Steve Kivelson and Xiao-Liang Qi that initiated this work and thanks Leon Balents, Mathew Fisher, Itamar Kimchi, Fa Wang, Ying Ran and Jason Alicea for helpful discussions. CMJ is supported by the David and Lucile Packard foundation and National Science Foundation under Grant No. NSF PHY11-25915. MPZ thanks Roger Mong for discussions.

\bibliography{TI}

\begin{thebibliography}{39}
\expandafter\ifx\csname natexlab\endcsname\relax\def\natexlab#1{#1}\fi
\expandafter\ifx\csname bibnamefont\endcsname\relax
  \def\bibnamefont#1{#1}\fi
\expandafter\ifx\csname bibfnamefont\endcsname\relax
  \def\bibfnamefont#1{#1}\fi
\expandafter\ifx\csname citenamefont\endcsname\relax
  \def\citenamefont#1{#1}\fi
\expandafter\ifx\csname url\endcsname\relax
  \def\url#1{\texttt{#1}}\fi
\expandafter\ifx\csname urlprefix\endcsname\relax\def\urlprefix{URL }\fi
\providecommand{\bibinfo}[2]{#2}
\providecommand{\eprint}[2][]{\url{#2}}

\bibitem[{\citenamefont{Lieb et~al.}(1961)\citenamefont{Lieb, Schultz, and
  Mattis}}]{Lieb1961}
\bibinfo{author}{\bibfnamefont{E.}~\bibnamefont{Lieb}},
  \bibinfo{author}{\bibfnamefont{T.}~\bibnamefont{Schultz}}, \bibnamefont{and}
  \bibinfo{author}{\bibfnamefont{D.}~\bibnamefont{Mattis}},
  \bibinfo{journal}{Annals of Physics} \textbf{\bibinfo{volume}{16}},
  \bibinfo{pages}{407 } (\bibinfo{year}{1961}), ISSN \bibinfo{issn}{0003-4916},
  \urlprefix\url{http://www.sciencedirect.com/science/article/pii/0003491661901154}.

\bibitem[{\citenamefont{Hastings}(2004)}]{Hastings2004}
\bibinfo{author}{\bibfnamefont{M.~B.} \bibnamefont{Hastings}},
  \bibinfo{journal}{Phys. Rev. B} \textbf{\bibinfo{volume}{69}},
  \bibinfo{pages}{104431} (\bibinfo{year}{2004}),
  \urlprefix\url{http://link.aps.org/doi/10.1103/PhysRevB.69.104431}.

\bibitem[{\citenamefont{Oshikawa}(2000)}]{Oshikawa}
\bibinfo{author}{\bibfnamefont{M.}~\bibnamefont{Oshikawa}},
  \bibinfo{journal}{Phys. Rev. Lett.} \textbf{\bibinfo{volume}{84}},
  \bibinfo{pages}{1535} (\bibinfo{year}{2000}),
  \urlprefix\url{http://link.aps.org/doi/10.1103/PhysRevLett.84.1535}.

\bibitem[{\citenamefont{Parameswaran et~al.}(2013)\citenamefont{Parameswaran,
  Turner, Arovas, and Vishwanath}}]{Parameswaran2013}
\bibinfo{author}{\bibfnamefont{S.~A.} \bibnamefont{Parameswaran}},
  \bibinfo{author}{\bibfnamefont{A.~M.} \bibnamefont{Turner}},
  \bibinfo{author}{\bibfnamefont{D.~P.} \bibnamefont{Arovas}},
  \bibnamefont{and}
  \bibinfo{author}{\bibfnamefont{A.}~\bibnamefont{Vishwanath}},
  \bibinfo{journal}{Nat Phys} \textbf{\bibinfo{volume}{9}},
  \bibinfo{pages}{299} (\bibinfo{year}{2013}),
  \urlprefix\url{http://dx.doi.org/10.1038/nphys2600}.

\bibitem[{\citenamefont{{Watanabe} et~al.}(2015)\citenamefont{{Watanabe}, {Po},
  {Vishwanath}, and {Zaletel}}}]{Watanabe2015}
\bibinfo{author}{\bibfnamefont{H.}~\bibnamefont{{Watanabe}}},
  \bibinfo{author}{\bibfnamefont{H.~C.} \bibnamefont{{Po}}},
  \bibinfo{author}{\bibfnamefont{A.}~\bibnamefont{{Vishwanath}}},
  \bibnamefont{and} \bibinfo{author}{\bibfnamefont{M.~P.}
  \bibnamefont{{Zaletel}}}, \bibinfo{journal}{ArXiv e-prints}
  (\bibinfo{year}{2015}), \eprint{1505.04193}.

\bibitem[{\citenamefont{Haldane}(1988)}]{Haldane1988}
\bibinfo{author}{\bibfnamefont{F.~D.~M.} \bibnamefont{Haldane}},
  \bibinfo{journal}{Phys. Rev. Lett.} \textbf{\bibinfo{volume}{61}},
  \bibinfo{pages}{1029} (\bibinfo{year}{1988}),
  \urlprefix\url{http://link.aps.org/doi/10.1103/PhysRevLett.61.1029}.

\bibitem[{\citenamefont{Read and Sachdev}(1990)}]{ReadSachdev1990}
\bibinfo{author}{\bibfnamefont{N.}~\bibnamefont{Read}} \bibnamefont{and}
  \bibinfo{author}{\bibfnamefont{S.}~\bibnamefont{Sachdev}},
  \bibinfo{journal}{Phys. Rev. B} \textbf{\bibinfo{volume}{42}},
  \bibinfo{pages}{4568} (\bibinfo{year}{1990}),
  \urlprefix\url{http://link.aps.org/doi/10.1103/PhysRevB.42.4568}.

\bibitem[{\citenamefont{Senthil et~al.}(2004)\citenamefont{Senthil, Vishwanath,
  Balents, Sachdev, and Fisher}}]{Senthil2004}
\bibinfo{author}{\bibfnamefont{T.}~\bibnamefont{Senthil}},
  \bibinfo{author}{\bibfnamefont{A.}~\bibnamefont{Vishwanath}},
  \bibinfo{author}{\bibfnamefont{L.}~\bibnamefont{Balents}},
  \bibinfo{author}{\bibfnamefont{S.}~\bibnamefont{Sachdev}}, \bibnamefont{and}
  \bibinfo{author}{\bibfnamefont{M.~P.~A.} \bibnamefont{Fisher}},
  \bibinfo{journal}{Science} \textbf{\bibinfo{volume}{303}},
  \bibinfo{pages}{1490} (\bibinfo{year}{2004}),
  \eprint{http://www.sciencemag.org/content/303/5663/1490.full.pdf},
  \urlprefix\url{http://www.sciencemag.org/content/303/5663/1490.abstract}.

\bibitem[{\citenamefont{Kimchi et~al.}(2013)\citenamefont{Kimchi, Parameswaran,
  Turner, Wang, and Vishwanath}}]{Kimchi2013}
\bibinfo{author}{\bibfnamefont{I.}~\bibnamefont{Kimchi}},
  \bibinfo{author}{\bibfnamefont{S.~A.} \bibnamefont{Parameswaran}},
  \bibinfo{author}{\bibfnamefont{A.~M.} \bibnamefont{Turner}},
  \bibinfo{author}{\bibfnamefont{F.}~\bibnamefont{Wang}}, \bibnamefont{and}
  \bibinfo{author}{\bibfnamefont{A.}~\bibnamefont{Vishwanath}},
  \bibinfo{journal}{Proceedings of the National Academy of Sciences}
  \textbf{\bibinfo{volume}{110}}, \bibinfo{pages}{16378}
  (\bibinfo{year}{2013}),
  \eprint{http://www.pnas.org/content/110/41/16378.full.pdf},
  \urlprefix\url{http://www.pnas.org/content/110/41/16378.abstract}.

\bibitem[{\citenamefont{{Ware} et~al.}(2015)\citenamefont{{Ware}, {Kimchi},
  {Parameswaran}, and {Bauer}}}]{BraydenBauer}
\bibinfo{author}{\bibfnamefont{B.}~\bibnamefont{{Ware}}},
  \bibinfo{author}{\bibfnamefont{I.}~\bibnamefont{{Kimchi}}},
  \bibinfo{author}{\bibfnamefont{S.~A.} \bibnamefont{{Parameswaran}}},
  \bibnamefont{and} \bibinfo{author}{\bibfnamefont{B.}~\bibnamefont{{Bauer}}},
  \bibinfo{journal}{ArXiv e-prints}  (\bibinfo{year}{2015}),
  \eprint{1507.00348}.

\bibitem[{\citenamefont{{Verstraete} and {Cirac}}(2004)}]{Verstraete2004}
\bibinfo{author}{\bibfnamefont{F.}~\bibnamefont{{Verstraete}}}
  \bibnamefont{and} \bibinfo{author}{\bibfnamefont{J.~I.}
  \bibnamefont{{Cirac}}}, \bibinfo{journal}{eprint arXiv:cond-mat/0407066}
  (\bibinfo{year}{2004}), \eprint{cond-mat/0407066}.

\bibitem[{\citenamefont{Pérez-García
  et~al.}(2010)\citenamefont{Pérez-García, Sanz, González-Guillén, Wolf,
  and Cirac}}]{García2010}
\bibinfo{author}{\bibfnamefont{D.}~\bibnamefont{Pérez-García}},
  \bibinfo{author}{\bibfnamefont{M.}~\bibnamefont{Sanz}},
  \bibinfo{author}{\bibfnamefont{C.~E.} \bibnamefont{González-Guillén}},
  \bibinfo{author}{\bibfnamefont{M.~M.} \bibnamefont{Wolf}}, \bibnamefont{and}
  \bibinfo{author}{\bibfnamefont{J.~I.} \bibnamefont{Cirac}},
  \bibinfo{journal}{New Journal of Physics} \textbf{\bibinfo{volume}{12}},
  \bibinfo{pages}{025010} (\bibinfo{year}{2010}),
  \urlprefix\url{http://stacks.iop.org/1367-2630/12/i=2/a=025010}.

\bibitem[{\citenamefont{Weichselbaum}(2012)}]{Weichselbaum2012}
\bibinfo{author}{\bibfnamefont{A.}~\bibnamefont{Weichselbaum}},
  \bibinfo{journal}{Annals of Physics} \textbf{\bibinfo{volume}{327}},
  \bibinfo{pages}{2972 } (\bibinfo{year}{2012}), ISSN
  \bibinfo{issn}{0003-4916},
  \urlprefix\url{http://www.sciencedirect.com/science/article/pii/S0003491612001121}.

\bibitem[{\citenamefont{Singh and Vidal}(2012)}]{Singh2012}
\bibinfo{author}{\bibfnamefont{S.}~\bibnamefont{Singh}} \bibnamefont{and}
  \bibinfo{author}{\bibfnamefont{G.}~\bibnamefont{Vidal}},
  \bibinfo{journal}{Phys. Rev. B} \textbf{\bibinfo{volume}{86}},
  \bibinfo{pages}{195114} (\bibinfo{year}{2012}),
  \urlprefix\url{http://link.aps.org/doi/10.1103/PhysRevB.86.195114}.

\bibitem[{\citenamefont{{Jiang} and {Ran}}(2015)}]{Jiang2015}
\bibinfo{author}{\bibfnamefont{S.}~\bibnamefont{{Jiang}}} \bibnamefont{and}
  \bibinfo{author}{\bibfnamefont{Y.}~\bibnamefont{{Ran}}},
  \bibinfo{journal}{ArXiv e-prints}  (\bibinfo{year}{2015}),
  \eprint{1505.03171}.

\bibitem[{\citenamefont{Singh et~al.}(2011)\citenamefont{Singh, Pfeifer, and
  Vidal}}]{Singh2011}
\bibinfo{author}{\bibfnamefont{S.}~\bibnamefont{Singh}},
  \bibinfo{author}{\bibfnamefont{R.~N.~C.} \bibnamefont{Pfeifer}},
  \bibnamefont{and} \bibinfo{author}{\bibfnamefont{G.}~\bibnamefont{Vidal}},
  \bibinfo{journal}{Phys. Rev. B} \textbf{\bibinfo{volume}{83}},
  \bibinfo{pages}{115125} (\bibinfo{year}{2011}),
  \urlprefix\url{http://link.aps.org/doi/10.1103/PhysRevB.83.115125}.

\bibitem[{\citenamefont{Bauer et~al.}(2011)\citenamefont{Bauer, Corboz, Or\'us,
  and Troyer}}]{Bauer2011}
\bibinfo{author}{\bibfnamefont{B.}~\bibnamefont{Bauer}},
  \bibinfo{author}{\bibfnamefont{P.}~\bibnamefont{Corboz}},
  \bibinfo{author}{\bibfnamefont{R.}~\bibnamefont{Or\'us}}, \bibnamefont{and}
  \bibinfo{author}{\bibfnamefont{M.}~\bibnamefont{Troyer}},
  \bibinfo{journal}{Phys. Rev. B} \textbf{\bibinfo{volume}{83}},
  \bibinfo{pages}{125106} (\bibinfo{year}{2011}),
  \urlprefix\url{http://link.aps.org/doi/10.1103/PhysRevB.83.125106}.

\bibitem[{\citenamefont{Singh and Vidal}(2013)}]{Singh2013}
\bibinfo{author}{\bibfnamefont{S.}~\bibnamefont{Singh}} \bibnamefont{and}
  \bibinfo{author}{\bibfnamefont{G.}~\bibnamefont{Vidal}},
  \bibinfo{journal}{Phys. Rev. B} \textbf{\bibinfo{volume}{88}},
  \bibinfo{pages}{121108} (\bibinfo{year}{2013}),
  \urlprefix\url{http://link.aps.org/doi/10.1103/PhysRevB.88.121108}.

\bibitem[{\citenamefont{{Williamson} et~al.}(2014)\citenamefont{{Williamson},
  {Bultinck}, {Mari{\"e}n}, {Sahinoglu}, {Haegeman}, and
  {Verstraete}}}]{Williamson2014}
\bibinfo{author}{\bibfnamefont{D.~J.} \bibnamefont{{Williamson}}},
  \bibinfo{author}{\bibfnamefont{N.}~\bibnamefont{{Bultinck}}},
  \bibinfo{author}{\bibfnamefont{M.}~\bibnamefont{{Mari{\"e}n}}},
  \bibinfo{author}{\bibfnamefont{M.~B.} \bibnamefont{{Sahinoglu}}},
  \bibinfo{author}{\bibfnamefont{J.}~\bibnamefont{{Haegeman}}},
  \bibnamefont{and}
  \bibinfo{author}{\bibfnamefont{F.}~\bibnamefont{{Verstraete}}},
  \bibinfo{journal}{ArXiv e-prints}  (\bibinfo{year}{2014}),
  \eprint{1412.5604}.

\bibitem[{\citenamefont{Haldane}(1983)}]{Haldane1983}
\bibinfo{author}{\bibfnamefont{F.}~\bibnamefont{Haldane}},
  \bibinfo{journal}{Physics Letters A} \textbf{\bibinfo{volume}{93}},
  \bibinfo{pages}{464 } (\bibinfo{year}{1983}), ISSN \bibinfo{issn}{0375-9601},
  \urlprefix\url{http://www.sciencedirect.com/science/article/pii/037596018390631X}.

\bibitem[{\citenamefont{Altland and Simons}(2010)}]{Altland_Simons2010}
\bibinfo{author}{\bibfnamefont{A.}~\bibnamefont{Altland}} \bibnamefont{and}
  \bibinfo{author}{\bibfnamefont{B.~D.} \bibnamefont{Simons}},
  \emph{\bibinfo{title}{Condensed Matter Field Theory, 2nd edition}}
  (\bibinfo{publisher}{Cambridge University Press}, \bibinfo{year}{2010}).

\bibitem[{\citenamefont{Ng}(1994)}]{Ng1994}
\bibinfo{author}{\bibfnamefont{T.-K.} \bibnamefont{Ng}},
  \bibinfo{journal}{Phys. Rev. B} \textbf{\bibinfo{volume}{50}},
  \bibinfo{pages}{555} (\bibinfo{year}{1994}),
  \urlprefix\url{http://link.aps.org/doi/10.1103/PhysRevB.50.555}.

\bibitem[{\citenamefont{Affleck et~al.}(1987)\citenamefont{Affleck, Kennedy,
  Lieb, and Tasaki}}]{AKLT}
\bibinfo{author}{\bibfnamefont{I.}~\bibnamefont{Affleck}},
  \bibinfo{author}{\bibfnamefont{T.}~\bibnamefont{Kennedy}},
  \bibinfo{author}{\bibfnamefont{E.~H.} \bibnamefont{Lieb}}, \bibnamefont{and}
  \bibinfo{author}{\bibfnamefont{H.}~\bibnamefont{Tasaki}},
  \bibinfo{journal}{Phys. Rev. Lett.} \textbf{\bibinfo{volume}{59}},
  \bibinfo{pages}{799} (\bibinfo{year}{1987}),
  \urlprefix\url{http://link.aps.org/doi/10.1103/PhysRevLett.59.799}.

\bibitem[{\citenamefont{Pollmann et~al.}(2012)\citenamefont{Pollmann, Berg,
  Turner, and Oshikawa}}]{Pollmann}
\bibinfo{author}{\bibfnamefont{F.}~\bibnamefont{Pollmann}},
  \bibinfo{author}{\bibfnamefont{E.}~\bibnamefont{Berg}},
  \bibinfo{author}{\bibfnamefont{A.~M.} \bibnamefont{Turner}},
  \bibnamefont{and} \bibinfo{author}{\bibfnamefont{M.}~\bibnamefont{Oshikawa}},
  \bibinfo{journal}{Phys. Rev. B} \textbf{\bibinfo{volume}{85}},
  \bibinfo{pages}{075125} (\bibinfo{year}{2012}),
  \urlprefix\url{http://link.aps.org/doi/10.1103/PhysRevB.85.075125}.

\bibitem[{\citenamefont{Fidkowski and Kitaev}(2011)}]{Fidkowski}
\bibinfo{author}{\bibfnamefont{L.}~\bibnamefont{Fidkowski}} \bibnamefont{and}
  \bibinfo{author}{\bibfnamefont{A.}~\bibnamefont{Kitaev}},
  \bibinfo{journal}{Phys. Rev. B} \textbf{\bibinfo{volume}{83}},
  \bibinfo{pages}{075103} (\bibinfo{year}{2011}),
  \urlprefix\url{http://link.aps.org/doi/10.1103/PhysRevB.83.075103}.

\bibitem[{\citenamefont{Chen et~al.}(2011)\citenamefont{Chen, Gu, and
  Wen}}]{Chen2011}
\bibinfo{author}{\bibfnamefont{X.}~\bibnamefont{Chen}},
  \bibinfo{author}{\bibfnamefont{Z.-C.} \bibnamefont{Gu}}, \bibnamefont{and}
  \bibinfo{author}{\bibfnamefont{X.-G.} \bibnamefont{Wen}},
  \bibinfo{journal}{Phys. Rev. B} \textbf{\bibinfo{volume}{83}},
  \bibinfo{pages}{035107} (\bibinfo{year}{2011}),
  \urlprefix\url{http://link.aps.org/doi/10.1103/PhysRevB.83.035107}.

\bibitem[{\citenamefont{Jiang et~al.}(2009)\citenamefont{Jiang, Kr\"uger,
  Moore, Sheng, Zaanen, and Weng}}]{Jiang2009}
\bibinfo{author}{\bibfnamefont{H.~C.} \bibnamefont{Jiang}},
  \bibinfo{author}{\bibfnamefont{F.}~\bibnamefont{Kr\"uger}},
  \bibinfo{author}{\bibfnamefont{J.~E.} \bibnamefont{Moore}},
  \bibinfo{author}{\bibfnamefont{D.~N.} \bibnamefont{Sheng}},
  \bibinfo{author}{\bibfnamefont{J.}~\bibnamefont{Zaanen}}, \bibnamefont{and}
  \bibinfo{author}{\bibfnamefont{Z.~Y.} \bibnamefont{Weng}},
  \bibinfo{journal}{Phys. Rev. B} \textbf{\bibinfo{volume}{79}},
  \bibinfo{pages}{174409} (\bibinfo{year}{2009}),
  \urlprefix\url{http://link.aps.org/doi/10.1103/PhysRevB.79.174409}.

\bibitem[{\citenamefont{{Wang} et~al.}(2015)\citenamefont{{Wang}, {Kivelson},
  and {Lee}}}]{Wang2015}
\bibinfo{author}{\bibfnamefont{F.}~\bibnamefont{{Wang}}},
  \bibinfo{author}{\bibfnamefont{S.}~\bibnamefont{{Kivelson}}},
  \bibnamefont{and} \bibinfo{author}{\bibfnamefont{D.-H.} \bibnamefont{{Lee}}},
  \bibinfo{journal}{ArXiv e-prints}  (\bibinfo{year}{2015}),
  \eprint{1501.00844}.

\bibitem[{\citenamefont{Freedman et~al.}(2004)\citenamefont{Freedman, Nayak,
  Shtengel, Walker, and Wang}}]{Freedman2004}
\bibinfo{author}{\bibfnamefont{M.}~\bibnamefont{Freedman}},
  \bibinfo{author}{\bibfnamefont{C.}~\bibnamefont{Nayak}},
  \bibinfo{author}{\bibfnamefont{K.}~\bibnamefont{Shtengel}},
  \bibinfo{author}{\bibfnamefont{K.}~\bibnamefont{Walker}}, \bibnamefont{and}
  \bibinfo{author}{\bibfnamefont{Z.}~\bibnamefont{Wang}},
  \bibinfo{journal}{Annals of Physics} \textbf{\bibinfo{volume}{310}},
  \bibinfo{pages}{428 } (\bibinfo{year}{2004}), ISSN \bibinfo{issn}{0003-4916},
  \urlprefix\url{http://www.sciencedirect.com/science/article/pii/S0003491604000260}.

\bibitem[{\citenamefont{Zhang et~al.}(2012)\citenamefont{Zhang, Grover, Turner,
  Oshikawa, and Vishwanath}}]{ZhangF2012}
\bibinfo{author}{\bibfnamefont{Y.}~\bibnamefont{Zhang}},
  \bibinfo{author}{\bibfnamefont{T.}~\bibnamefont{Grover}},
  \bibinfo{author}{\bibfnamefont{A.}~\bibnamefont{Turner}},
  \bibinfo{author}{\bibfnamefont{M.}~\bibnamefont{Oshikawa}}, \bibnamefont{and}
  \bibinfo{author}{\bibfnamefont{A.}~\bibnamefont{Vishwanath}},
  \bibinfo{journal}{Phys. Rev. B} \textbf{\bibinfo{volume}{85}},
  \bibinfo{pages}{235151} (\bibinfo{year}{2012}),
  \urlprefix\url{http://link.aps.org/doi/10.1103/PhysRevB.85.235151}.

\bibitem[{\citenamefont{Kitaev and Preskill}(2006)}]{Kitaev2006EE}
\bibinfo{author}{\bibfnamefont{A.}~\bibnamefont{Kitaev}} \bibnamefont{and}
  \bibinfo{author}{\bibfnamefont{J.}~\bibnamefont{Preskill}},
  \bibinfo{journal}{Phys. Rev. Lett.} \textbf{\bibinfo{volume}{96}},
  \bibinfo{pages}{110404} (\bibinfo{year}{2006}),
  \urlprefix\url{http://link.aps.org/doi/10.1103/PhysRevLett.96.110404}.

\bibitem[{\citenamefont{Levin and Wen}(2006)}]{Levin2006EE}
\bibinfo{author}{\bibfnamefont{M.}~\bibnamefont{Levin}} \bibnamefont{and}
  \bibinfo{author}{\bibfnamefont{X.-G.} \bibnamefont{Wen}},
  \bibinfo{journal}{Phys. Rev. Lett.} \textbf{\bibinfo{volume}{96}},
  \bibinfo{pages}{110405} (\bibinfo{year}{2006}),
  \urlprefix\url{http://link.aps.org/doi/10.1103/PhysRevLett.96.110405}.

\bibitem[{\citenamefont{Zhao et~al.}(2012)\citenamefont{Zhao, Xu, Chen, Wei,
  Qin, Zhang, and Xiang}}]{Zhao2012}
\bibinfo{author}{\bibfnamefont{H.~H.} \bibnamefont{Zhao}},
  \bibinfo{author}{\bibfnamefont{C.}~\bibnamefont{Xu}},
  \bibinfo{author}{\bibfnamefont{Q.~N.} \bibnamefont{Chen}},
  \bibinfo{author}{\bibfnamefont{Z.~C.} \bibnamefont{Wei}},
  \bibinfo{author}{\bibfnamefont{M.~P.} \bibnamefont{Qin}},
  \bibinfo{author}{\bibfnamefont{G.~M.} \bibnamefont{Zhang}}, \bibnamefont{and}
  \bibinfo{author}{\bibfnamefont{T.}~\bibnamefont{Xiang}},
  \bibinfo{journal}{Phys. Rev. B} \textbf{\bibinfo{volume}{85}},
  \bibinfo{pages}{134416} (\bibinfo{year}{2012}),
  \urlprefix\url{http://link.aps.org/doi/10.1103/PhysRevB.85.134416}.

\bibitem[{\citenamefont{Affleck et~al.}(1988)\citenamefont{Affleck, Kennedy,
  Lieb, and Tasaki}}]{AKLT1988}
\bibinfo{author}{\bibfnamefont{I.}~\bibnamefont{Affleck}},
  \bibinfo{author}{\bibfnamefont{T.}~\bibnamefont{Kennedy}},
  \bibinfo{author}{\bibfnamefont{E.}~\bibnamefont{Lieb}}, \bibnamefont{and}
  \bibinfo{author}{\bibfnamefont{H.}~\bibnamefont{Tasaki}},
  \bibinfo{journal}{Communications in Mathematical Physics}
  \textbf{\bibinfo{volume}{115}}, \bibinfo{pages}{477} (\bibinfo{year}{1988}),
  ISSN \bibinfo{issn}{0010-3616},
  \urlprefix\url{http://dx.doi.org/10.1007/BF01218021}.

\bibitem[{\citenamefont{Vidal}(2007)}]{Vidal2007}
\bibinfo{author}{\bibfnamefont{G.}~\bibnamefont{Vidal}},
  \bibinfo{journal}{Phys. Rev. Lett.} \textbf{\bibinfo{volume}{98}},
  \bibinfo{pages}{070201} (\bibinfo{year}{2007}),
  \urlprefix\url{http://link.aps.org/doi/10.1103/PhysRevLett.98.070201}.

\bibitem[{\citenamefont{Or\'us and Vidal}(2008)}]{Orus2008}
\bibinfo{author}{\bibfnamefont{R.}~\bibnamefont{Or\'us}} \bibnamefont{and}
  \bibinfo{author}{\bibfnamefont{G.}~\bibnamefont{Vidal}},
  \bibinfo{journal}{Phys. Rev. B} \textbf{\bibinfo{volume}{78}},
  \bibinfo{pages}{155117} (\bibinfo{year}{2008}),
  \urlprefix\url{http://link.aps.org/doi/10.1103/PhysRevB.78.155117}.

\bibitem[{\citenamefont{Wei et~al.}(2011)\citenamefont{Wei, Affleck, and
  Raussendorf}}]{Wei2011}
\bibinfo{author}{\bibfnamefont{T.-C.} \bibnamefont{Wei}},
  \bibinfo{author}{\bibfnamefont{I.}~\bibnamefont{Affleck}}, \bibnamefont{and}
  \bibinfo{author}{\bibfnamefont{R.}~\bibnamefont{Raussendorf}},
  \bibinfo{journal}{Phys. Rev. Lett.} \textbf{\bibinfo{volume}{106}},
  \bibinfo{pages}{070501} (\bibinfo{year}{2011}),
  \urlprefix\url{http://link.aps.org/doi/10.1103/PhysRevLett.106.070501}.

\bibitem[{\citenamefont{Yao and Kivelson}(2010)}]{YaoKivelson}
\bibinfo{author}{\bibfnamefont{H.}~\bibnamefont{Yao}} \bibnamefont{and}
  \bibinfo{author}{\bibfnamefont{S.~A.} \bibnamefont{Kivelson}},
  \bibinfo{journal}{Phys. Rev. Lett.} \textbf{\bibinfo{volume}{105}},
  \bibinfo{pages}{166402} (\bibinfo{year}{2010}),
  \urlprefix\url{http://link.aps.org/doi/10.1103/PhysRevLett.105.166402}.

\bibitem[{\citenamefont{Perez-Garcia et~al.}(2008)\citenamefont{Perez-Garcia,
  Verstraete, Cirac, and Wolf}}]{Perez-Garcia2008}
\bibinfo{author}{\bibfnamefont{D.}~\bibnamefont{Perez-Garcia}},
  \bibinfo{author}{\bibfnamefont{F.}~\bibnamefont{Verstraete}},
  \bibinfo{author}{\bibfnamefont{J.~I.} \bibnamefont{Cirac}}, \bibnamefont{and}
  \bibinfo{author}{\bibfnamefont{M.~M.} \bibnamefont{Wolf}},
  \bibinfo{journal}{Quant. Inf. Comp.} \textbf{\bibinfo{volume}{8}},
  \bibinfo{pages}{0650} (\bibinfo{year}{2008}).

\end{thebibliography}
\end{document}